 \documentclass[review,3p,times]{elsarticle}
\usepackage{amsmath,hyperref}
\usepackage{lineno}
\usepackage{color}
\usepackage{subfigure}
\usepackage{epsfig}
\usepackage{caption,xcolor,float,graphicx}
\usepackage{graphics}
\usepackage{epstopdf}
\usepackage{soul}
\usepackage{tikz}
\usepackage{xcolor}
\usepackage{booktabs}
\usepackage{array}

\definecolor{rainbow1}{RGB}{210,0,0}
\definecolor{rainbow2}{RGB}{255,127,0}
\definecolor{rainbow3}{RGB}{255,255,85}
\definecolor{rainbow4}{RGB}{0,210,0}
\definecolor{rainbow5}{RGB}{85,255,255}
\definecolor{rainbow6}{RGB}{56,56,255}
\definecolor{rainbow7}{RGB}{170,0,170}

\journal{.}

\begin{document}

\begin{frontmatter}
	
\title{A fluctuating lattice Boltzmann method for viscoelastic fluid flows}
\author[mymainaddress]{Juanyong Wang}
\address[mymainaddress]{School of Mathematics and Physics, China University of Geosciences, Wuhan 430074, China}
\author[mysecondaddress]{Xinyue Liu}
\address[mysecondaddress]{Department of Mathematics, College of Science, National University of Defense Technology, Changsha, 410073, China}
\author[mymainaddress]{Lei Wang\corref{mycorrespondingauthor}}
\cortext[mycorrespondingauthor]{Corresponding author}
\ead{leiwang@cug.edu.cn}

\author[mythirdaddress]{Yuan Yu}
\address[mythirdaddress]{School of Mathematics and Computational Science, Xiangtan University, Xiangtan, 411105, China}

\author[mymainaddress]{Yiran Ji}

\begin{abstract}

This study introduces a novel fluctuating lattice Boltzmann (LB) method for simulating viscoelastic fluid flows governed by the Oldroyd-B model. In contrast to conventional LB approaches that explicitly compute the divergence of the polymer stress tensor using finite-difference schemes, the proposed method incorporates the polymer stress implicitly by introducing a polymer stress fluctuation term directly into the evolution equation. This treatment avoids the need for stress-gradient computations, and preserves the physical characteristics of viscoelastic fluid flows. The proposed method is validated against four classical benchmark problems: the simplified four-roll mill, planar Poiseuille flow, unsteady Womersley flow, and the three-dimensional Taylor–Green vortex. The numerical results show excellent agreement with analytical solutions and previous numerical results, confirming the method’s reliability in viscoelastic fluid dynamics. Moreover, performance evaluations demonstrate that the present model reduces the memory occupancy and enhances computational efficiency, highlighting its potential for large-scale simulations of complex viscoelastic flows systems.

\end{abstract}
	
\end{frontmatter}

\section{Introduction}\label{section_1}

Viscoelastic fluid represents a distinct non-Newtonian fluid class that simultaneously exhibits viscous and elastic characteristics. Its complex flow behavior arises in scientific and industrial contexts, including advanced manufacturing processes such as 3D printing \cite{Skylar_NATURE2019}, the formulation of medical hemostatic gels \cite{Mao_JMST2021}, the deployment of drilling fluids in subsurface exploration \cite{Karakosta_JMS2021}, and the utilization of drag-reducing agents in pipeline transport systems \cite{Karami_JNNFM2012}. Unlike Newtonian fluids, viscoelastic fluid is typically composed of long-chain polymer molecules or other structurally intricate constituents, which endow the material with memory effects. Consequently, the stress at any given moment depends not only on the current deformation but also on the entire deformation history experienced by the fluid element \cite{Denn_ARFM1990}. This memory-dependent behavior gives rise to a range of counterintuitive and uniquely viscoelastic phenomena, including the rod climbing effect, known as the Weissenberg effect \cite{More_SoftMatter2023}, and the siphoning phenomena without channel \cite{James_NATURE1966}. These phenomena demonstrate the intrinsic complexity of viscoelastic flows and highlight the need for accurate theoretical and numerical frameworks to capture their complex dynamics.

Numerical simulation has emerged as a powerful tool in studying viscoelastic fluid flows, offering deep insights into their complex flow manners \cite{Alves_ARFM2021}. In contrast to Newtonian fluid simulations that primarily involve the Navier–Stokes equations (i.e., the mass and momentum conservation equations), simulations of viscoelastic fluid flows require additional constitutive equations or rheological models to describe the relationship between polymer stress and the material's deformation rate \cite{Alves_ARFM2021}. Differential constitutive equations, which are widely used and derived from the framework of continuum mechanics, can effectively model the elastic stretching and relaxation behavior of polymer chains \cite{Bird_ARFM1995}. Differential constitutive equations rooted in continuum mechanics are widely used, with notable examples including the Oldroyd-B model \cite{Oldroyd_SAMPS1950}, the Giesekus model \cite{Giesekus_JNNFM1982}, and the Phan-Thien–Tanner (PTT) model \cite{Thien_JNNFM1977}. To solve the highly coupled and nonlinear governing equations, numerous conventional macroscopic numerical methods have been developed \cite{Kimura_JSC2019,Crochet_RCT1989,Lee_JSC2019,Aboubacar_JCP2004}. Tomé et al. \cite{Tome_JCP2008} developed a finite difference framework incorporating integral constitutive models to simulate viscoelastic flows. Their approach successfully reproduced key experimental features of planar contraction flows, including the formation of lip vortices and pronounced stress concentrations near the re-entrant corner. Moreno et al. \cite{Moreno_JNNFM2021} utilized the finite element method to simulate viscoelastic fluid flows under non-isothermal conditions. The results demonstrated that the stress in the flow around a cylinder under non-isothermal conditions was reduced by more than 30\% compared to isothermal conditions. In contrast to conventional macroscopic numerical methods, the lattice Boltzmann (LB) method, as a mesoscopic method grounded in kinetic theory, offers distinct advantages in simulating complex hydrodynamic phenomena \cite{Aidun_ARFM2010,Li_PECS2016}. Its inherently modular structure, high parallel efficiency, and ease in incorporating complex boundary conditions and multi-physics interactions have made the LB method a powerful and flexible framework for capturing the intricate behavior of non-Newtonian fluids using in various applications \cite{Phillips_JAM2011,Yoshino_JNNFM2007}.

Over the past few decades, the LB method has been successfully applied in viscoelastic fluid, and research results have shown that this method can effectively capture the motion characteristics of viscoelastic fluid flows \cite{Osmanlic_CF2016,Dzanic_CF2022}. Qian et al. \cite{Qian_PRL1997} successfully incorporated elastic effects into viscoelastic materials by modifying the equilibrium distribution function. Similarly, Onishi et al. \cite{Onishi_PhysicaA2006} incorporated a viscoelastic stress term into the equilibrium distribution function, effectively simulating the non-Newtonian behavior of viscoelastic fluids. With this modification, the stress of the fluid is determined by both the viscous stress and the viscoelastic stress. Malaspinas et al. \cite{Malaspinas_JNNFM2010} solved the Oldroyd-B constitutive equation using a modified “advection-diffusion” scheme and achieved accurate reproduction of the results obtained using a high-accuracy spectral method in the Taylor-Green vortex benchmark. Subsequently, Su et al. \cite{Su_JNNFM2013} proposed a new numerical scheme using the lattice Boltzmann method to simulate the viscoelastic fluid flows at high Weissenberg numbers. Their results indicate that the model is stable under these conditions and effectively captures the structural patterns associated with elastic instability.

Although considerable progress has been made in simulating viscoelastic fluid flows using the LB method, several critical challenges remain unresolved. Most existing studies focus on solving the polymer stress field, wherein the divergence of the polymer stress tensor is incorporated as an external body force in the momentum equation \cite{Malaspinas_JNNFM2010,Su_JNNFM2013}. While this approach enables coupling between the flow and the polymer stress, it also introduces potential numerical and computational difficulties. Specifically, the evaluation of the stress tensor derivatives requires additional finite-difference schemes, which not only increase algorithmic complexity but also amplify the potential for numerical error propagation. In addition, the explicit storage and evolution of the whole stress tensor field impose a substantial memory burden, constraining the utilization of the LB method for three-dimensional large-scale simulations.

To address these limitations, the present work draws inspiration from a fluctuating LB method framework developed by Nie et al. \cite{Nie_Particuology2009}, and proposes a novel fluctuating LB method for viscoelastic fluid flows governed by the Oldroyd-B model. By introducing a polymer stress fluctuation term directly into the evolution equation, the influence of polymer stress is incorporated implicitly, thereby avoiding the need to compute the stress divergence explicitly. This treatment not only preserves the physical characteristics of viscoelastic fluid flows but also enhances computational efficiency and releases memory burden. The outline of the paper is as follows. Section \ref{section_2} describes the mathematical method used. The numerical examples, which included four tests, are presented in Section \ref{section_3}, followed by a brief conclusion in Section \ref{section_4}.

\section{Mathematical method}\label{section_2}

\subsection{Governing equations}\label{section_2.1}

In the present work, we focus on the macroscopic governing equations for isothermal and incompressible viscoelastic fluid flows, which are characterized by the Oldroyd-B constitutive equation. This model is derived from the kinetic theory of polymer dilute solutions and is widely used to describe the elastic behavior of viscoelastic fluid \cite{Su_PRE2013}. The Oldroyd-B model predicts a quadratic first normal stress difference and constant viscosity, making it a reliable benchmark for testing numerical algorithms \cite{Su_JNNFM2013}. In our simulations, we take the incompressible flow as a solvent model, and the governing equations of the incompressible mass and momentum conservations can be expressed as
\begin{equation}\label{mass_conservation}
	\nabla  \cdot {\bf{u}} = 0,
\end{equation}
\begin{equation}\label{momentum_conservation}
	\frac{{\partial {\bf{u}}}}{{\partial t}} + \nabla  \cdot {\bf{uu}} =  - \nabla  \cdot p{\bf{I}} + \nabla  \cdot \tau_s + {\nabla  \cdot \tau_p} + {\bf{F}},
\end{equation}
in which $\bf{u}$ represents the velocity, $p$ denotes the pressure, and ${\bf{F}}$ is the body force. $\tau_s$ and $\tau_p$ denote the solvent stress and the polymer stress, respectively. $\tau_s$ is given by
\begin{equation}
	\tau_s=\left[ {{\mu _s}\left( {\nabla {\bf{u}} + {{ {\nabla {\bf{u}}} }^{\rm{T}}}} \right)} \right],
\end{equation}
where ${\mu _s}$ represents the solvent viscosity. $\tau_p$ can be expressed via a strain function as
\begin{equation}
	\tau_p=\left[ {\frac{{{\mu _p}}}{\lambda }\left( {{\bf{A}} - {\bf{I}}} \right)} \right],
\end{equation}
in which $\mu_p$ is the dynamic viscosity of polymer, $\lambda$ denotes the relaxation time of the polymer, and $\bf{A}$ represents the conformation tensor. The Oldroyd-B constitutive equation for viscoelastic fluid can be written as
\begin{equation}\label{constitutive_equation}
	\frac{{\partial {\bf{A}}}}{{\partial t}} + {\bf{u}} \cdot \nabla {\bf{A}} =  - \frac{1}{\lambda }\left( {{\bf{A}} - {\bf{I}}} \right) + {\bf{A}} \cdot \nabla {\bf{u}} + {\left( {\nabla {\bf{u}}} \right)^{\rm{T}}} \cdot {\bf{A}}.
\end{equation}

\subsection{Fluctuating lattice Boltzmann model for the velocity field}

To solve the equations of the incompressible mass and momentum conservations, we employ the LB Method, a highly effective and versatile approach for solving incompressible and weakly compressible Navier–Stokes equations. The method's inherent simplicity and computational efficiency make it particularly suitable for simulating complex fluid dynamics with reduced computational overhead and enhanced numerical stability. Generally, the divergence of polymer stress in the momentum equation is incorporated as an source term in the momentum equation. The present work draws inspiration from a fluctuating LB method framework developed by Nie et al. \cite{Nie_Particuology2009}, and proposes a novel fluctuating LB method to characterize the effect of polymer stress. Specifically, our approach introduces a polymer stress fluctuation term into the evolution equation, thereby eliminating the need to calculate the polymer stress tensor's divergence. The evolution equation in discrete velocity space can be written as
\begin{equation}\label{evolution_equation}
	{f_i}\left( {{\bf{x}} + {{\bf{c}}_i}\Delta t,t + \Delta t} \right) - {f_i}({\bf{x}},t) =  - \frac{1}{{{\tau _f}}}\left[ {{f_i}({\bf{x}},t) - f_i^{({\rm{eq}})}({\bf{x}},t)} \right] + f_i^B\left( {{\bf{x}},t} \right) + \Delta t{F_i}\left( {{\bf{x}},t} \right),
\end{equation}
in which $f_i$ is the distribution function at position $\bf{x}$ and time $t$, $\tau_f$ denotes the dimensionless relaxation time, and $f_i^B$ represents the polymer stress fluctuation term. $F_i$ is the discrete forcing term accounting for the external force, and it is defined as 
\begin{equation}
	{F_i}\left( {{\bf{x}},t} \right) = {\omega _i}\left( {1 - \frac{1}{{2{\tau _f}}}} \right)\left[ {\frac{{{{\bf{c}}_i} \cdot {\bf{F}}}}{{c_s^2}} + \frac{{({\bf{Fu}} + {\bf{uF}}):\left( {{{\bf{c}}_i}{{\bf{c}}_i} - c_s^2{\bf{I}}} \right)}}{{2c_s^4}}} \right].
\end{equation}
$f_i^{({\rm{eq}})}$ is the local equilibrium distribution function, and it is given by 
\begin{equation}
	f_i^{(\rm{eq})}({\bf{x}},t) = {\lambda _i}p + {\omega _i}\left[ {\frac{{{{\bf{c}}_i} \cdot {\bf{u}}}}{{c_s^2}} + \frac{{{\bf{uu}}:\left( {{{\bf{c}}_i}{{\bf{c}}_i} - c_s^2{\bf{I}}} \right)}}{{2c_s^4}}} \right],
\end{equation}
where $p$ is the pressure, ${\lambda _i}={\omega_i}/c_s^{2}$, ${\lambda _0}=({\omega_0}-1)/c_s^{2}$, $c_s^{2}=c^{2}/3$. The fluid velocity is computed by                          
\begin{equation}\label{u_macro}
	{\bf{u}} = \sum\limits_{i \ne 0} {{{\bf{c}}_i}f}  + \frac{{\Delta t}}{2}{\bf{F}},
\end{equation}
and the pressure of flow can be calculated as (see more details in Appendix B.)
\begin{equation}
	p = \frac{{c_s^2}}{{1 - {\omega _0}}}\left[ {\sum\limits_{i \ne 0} {{f_i}}  + {s_0}\left( {\bf{u}} \right) + {\tau _f}\Delta t{F_0} + {\tau _f}f_0^B} \right],
\end{equation}
in which
\begin{equation}
	s_i(\mathbf{u})=\omega_i\left[3 \frac{\left(\mathbf{e}_i \cdot \mathbf{u}\right)}{c}+4.5 \frac{\left(\mathbf{e}_i \cdot \mathbf{u}\right)^2}{c^2}-1.5 \frac{|\mathbf{u}|^2}{c^2}\right].
\end{equation}
For the two-dimensional case considered here, the classical $\rm{D2Q9}$ model is employed in the present work, and the discrete velocities $\mathbf{c}_i$ is given by
\begin{equation}
	\mathbf{c}_i=c \left[\begin{array}{ccccccccc}
		0 & 1 & 0 & -1 & 0 & 1 & -1 & -1 & 1 \\
		0 & 0 & 1 & 0 & -1 & 1 & 1 & -1 & -1
	\end{array}\right],
\end{equation}
where $c={\Delta x}/{\Delta t}$ represents the lattice velocity, ${\Delta x}$ and ${\Delta t}$ denote the lattice spacing and time step, respectively. The corresponding weight coefficients are
\begin{equation}
	{\omega _i} = \left\{ {\begin{array}{*{20}{c}}
			{\frac{4}{9}}&{i = 0,}\\
			{\frac{1}{9}}&{i = 1,2,3,4,}\\
			{\frac{1}{{36}}}&{i = 5,6,7,8}.
	\end{array}} \right.
\end{equation}
By the Chapman-Enskog expansion analysis, the Eq.(\ref{evolution_equation}) can recover the incompressible mass and momentum conservations (i.e. Eq.(\ref{mass_conservation}) and Eq.(\ref{momentum_conservation})) with
\begin{equation}
	{\tau _f} = \frac{{{\mu _s}}}{{c_s^2\Delta t}} + 0.5.
\end{equation}
By employing multi-scale analysis with an expansion parameter $\epsilon$, we have
\begin{subequations}
	\begin{gather}
		{f_i} = f_i^{\left( 0 \right)} + \varepsilon f_i^{\left( 1 \right)} + {\varepsilon ^2}f_i^{\left( 2 \right)} +  \cdot  \cdot  \cdot {\rm{, }}  
		f_i^B = \varepsilon f_i^{\left( {B1} \right)} , \label{f_expansion} \\
		\frac{\partial }{{\partial t}} = \varepsilon \frac{\partial }{{\partial {t_1}}} + {\varepsilon ^2}\frac{\partial }{{\partial {t_2}}}, 
		{\rm{ }}\nabla  = \varepsilon {\nabla _1}{\rm{, }} 
		{F_i} = \varepsilon F_i^{\left( 1 \right)}, 
	\end{gather}
\end{subequations}
in which ${\partial  \mathord{\left/{\vphantom {\partial  {\partial t}}} \right.\kern-\nulldelimiterspace} {\partial t}}$ and $\nabla$ are the temporal and spatial derivative, respectively. Additionally, $f_i^{(eq)}$, $f_i$ and $F_i$ satisfy the following conservation laws
\begin{subequations}\label{moments_condition}
	\begin{gather}
		\sum\limits_{i} {f_i^{\left( {{\rm{eq}}} \right)}}  = \sum\limits_{i} {{f_i}}  = 0{\rm{, }}
		\sum\limits_{i} {{{\bf{c}}_i}f_i^{\left( {{\rm{eq}}} \right)}}  = {\bf{u}}{\rm{, }}\label{Aa}\\
		\sum\limits_{i} {{{\bf{c}}_i}{{\bf{c}}_i}f_i^{\left( {{\rm{eq}}} \right)}}  = {\bf{uu}} + p{\bf{I}}{\rm{, }}
		\sum\limits_{i} {{{\bf{c}}_i}{{\bf{c}}_i}{{\bf{c}}_i}f_i^{\left( {{\rm{eq}}} \right)}}  = c_s^2\left( {{\delta _{\alpha \beta }}{{\bf{u}}_\gamma } + {\delta _{\alpha \gamma }}{{\bf{u}}_\beta } + {\delta _{\beta \gamma }}{{\bf{u}}_\alpha }} \right),\label{Ab}\\
		\sum\limits_{i} {{F_i}}  = 0{\rm{, }}
		\sum\limits_{i} {{{\bf{c}}_i}{F_i}}  = \left( {1 - \frac{1}{{2{\tau _f}}}} \right){\bf{F}}{\rm{, }}
		\sum\limits_{i} {{{\bf{c}}_i}{{\bf{c}}_i}{F_i}}  = \left( {1 - \frac{1}{{2{\tau _f}}}} \right)\left( {{\bf{Fu}} + {\bf{uF}}} \right),\label{Ac}
	\end{gather}
\end{subequations}
where ${{\delta _{\alpha \beta }}}$ is the Kronecker tensor. Performing a Taylor series expansion of Eq.(\ref{evolution_equation}) to second-order and substituting the above multi-scale expansion, we obtain the zeroth-order, first-order, and second-order equations in $\epsilon$, respectively, as follows
\begin{subequations}
	\begin{gather}
		O\left( {{\varepsilon ^0}} \right):{\rm{ }}f_i^{\left( 0 \right)} = f_i^{\left( {{\rm{eq}}} \right)},\label{epsilon_0}\\
		O\left( {{\varepsilon ^1}} \right):{\rm{ }}{D_{1i}}f_i^{\left( 0 \right)} =  - \frac{1}{{{\tau _f}\Delta t}}f_i^{\left( 1 \right)} + \frac{1}{{\Delta t}}f_i^{\left( {B1} \right)} + F_i^{(1)},\label{epsilon_1}\\
		O\left( {{\varepsilon ^2}} \right):{\rm{ }}{D_{1i}}f_i^{\left( 1 \right)} + \frac{\partial }{{\partial {t_2}}}f_i^{\left( 0 \right)} + \frac{{\Delta t}}{2}D_{1i}^2f_i^{\left( 0 \right)} =  - \frac{1}{{{\tau _f}\Delta t}}f_i^{\left( 2 \right)},\label{epsilon_2}
	\end{gather}
\end{subequations}
in which ${D_{1i}} = {\partial  \mathord{\left/{\vphantom {\partial  {\partial {t_1}}}} \right.\kern-\nulldelimiterspace} {\partial {t_1}}} + {{\bf{c}}_i}{\nabla _1}$. Taking the zero- and first-order moments of Eq.(\ref{f_expansion}) and combining Eq.(\ref{u_macro}), we have
\begin{equation}\label{epsilon_0_moments}
	\sum\limits_{i} {f_i^{\left( k \right)}}  = 0,{\rm{  }}k = 1,2,{\rm{ }}\sum\limits_{i} {{{\bf{c}}_i}f_i^{\left( 1 \right)}}  =  - \frac{{\Delta t}}{2}{{\bf{F}}^{\left( 1 \right)}},{\rm{  }}\sum\limits_{i} {{{\bf{c}}_i}f_i^{\left( 2 \right)}}  = 0.
\end{equation}
According to Eq.(\ref{epsilon_0_moments}), we obtain the moments of Eq.(\ref{epsilon_1})
\begin{equation}\label{zero_order_epsilon_1}
	{\nabla _1} \cdot {\bf{u}} = \frac{1}{{\Delta t}}\sum\limits_{i} {f_i^{\left( {B1} \right)}} ,
\end{equation}
\begin{equation}\label{first_order_epsilon_1}
	\frac{\partial }{{\partial {t_1}}}{\bf{u}} + {\nabla _1} \cdot \left( {{\bf{uu}} + p{\bf{I}}} \right) = \frac{1}{{\Delta t}}\sum\limits_{i} {{{\bf{c}}_i}f_i^{\left( {B1} \right)}}  + {{\bf{F}}^{\left( 1 \right)}}.
\end{equation}
Moreover, applying Eq.(\ref{epsilon_1}) to Eq.(\ref{epsilon_2}), one can obtain
\begin{equation}\label{epsilon_2_expansion}
	\frac{\partial }{{\partial {t_2}}}f_i^{\left( 0 \right)} + {D_{1i}}\left( {1 - \frac{1}{{2{\tau _f}}}} \right)f_i^{\left( 1 \right)} + \frac{1}{2}{D_{1i}}f_i^{\left( {B1} \right)} + \frac{{\Delta t}}{2}{D_{1i}}F_i^{\left( 1 \right)} =  - \frac{1}{{{\tau _f}\Delta t}}f_i^{\left( 2 \right)}.
\end{equation}
Summing Eq.(\ref{epsilon_2_expansion}) over $i$, we get the scale equation as
\begin{equation}\label{zero_order_epsilon_2}
	\frac{\partial }{{\partial {t_1}}}\sum\limits_{i} {f_i^{\left( {B1} \right)}}  + {\nabla _1} \cdot \sum\limits_{i} {{{\bf{c}}_i}f_i^{\left( {B1} \right)}}  = 0.
\end{equation}
Taking the first-order moments of Eq.(\ref{epsilon_2_expansion}), we get
\begin{equation}\label{first_order_epsilon_2}
	\frac{\partial }{{\partial {t_2}}}{\bf{u}} + \left( {1 - \frac{1}{{2{\tau _f}}}} \right){\nabla _1} \cdot \sum\limits_{i} {{{\bf{c}}_i}{{\bf{c}}_i}f_i^{\left( 1 \right)}}  + \frac{1}{2}\left( {\frac{\partial }{{\partial {t_1}}}\sum\limits_{i} {{{\bf{c}}_i}f_i^{\left( {B1} \right)}}  + {\nabla _1} \cdot \sum\limits_{i} {{{\bf{c}}_i}{{\bf{c}}_i}f_i^{\left( {B1} \right)}} } \right) + \frac{{\Delta t}}{2}{\nabla _1} \cdot \sum\limits_{i} {{{\bf{c}}_i}{{\bf{c}}_i}F_i^{\left( 1 \right)}} ,
\end{equation}
in which the second-order moments of $f_i^{\left( 1 \right)}$ can be derived from Eq.(\ref{epsilon_1}) as
\begin{equation}\label{second_order_f1}
	\sum\limits_{i} {{{\bf{c}}_i}{{\bf{c}}_i}f_i^{\left( 1 \right)}}  =  - {\tau _f}\Delta t\left( {\frac{\partial }{{\partial {t_1}}}\sum\limits_{i} {{{\bf{c}}_i}{{\bf{c}}_i}f_i^{\left( 0 \right)}}  + {\nabla _1} \cdot \sum\limits_{i} {{{\bf{c}}_i}{{\bf{c}}_i}{{\bf{c}}_i}f_i^{\left( 0 \right)}}  - \frac{1}{{\Delta t}}\sum\limits_{i} {{{\bf{c}}_i}{{\bf{c}}_i}f_i^{\left( {B1} \right)}}  - \sum\limits_{i} {{{\bf{c}}_i}{{\bf{c}}_i}F_i^{\left( 1 \right)}} } \right),
\end{equation}
then substituting Eq.(\ref{moments_condition}) into Eq.(\ref{second_order_f1}), we have
\begin{equation}
\begin{aligned}
		\sum\limits_{i} {{{\bf{c}}_i}{{\bf{c}}_i}f_i^{\left( 1 \right)}}  &=  - {\tau _f}\Delta t\left( {\frac{\partial }{{\partial {t_1}}}\sum\limits_{i} {{{\bf{c}}_i}{{\bf{c}}_i}f_i^{\left( 0 \right)}}  + {\nabla _1} \cdot \sum\limits_{i} {{{\bf{c}}_i}{{\bf{c}}_i}{{\bf{c}}_i}f_i^{\left( 0 \right)}}  - \frac{1}{{\Delta t}}\sum\limits_{i} {{{\bf{c}}_i}{{\bf{c}}_i}f_i^{\left( {B1} \right)}}  - \sum\limits_{i} {{{\bf{c}}_i}{{\bf{c}}_i}F_i^{\left( 1 \right)}} } \right)\\&
		=  - {\tau _f}\Delta t\left[ {\frac{\partial }{{\partial {t_1}}}\left( {{\bf{uu}} + p{\bf{I}}} \right) + c_s^2{\nabla _1} \cdot \left( {{\delta _{\alpha \beta }}{{\bf{u}}_\gamma } + {\delta _{\alpha \gamma }}{{\bf{u}}_\beta } + {\delta _{\beta \gamma }}{{\bf{u}}_\alpha }} \right) - \frac{1}{{\Delta t}}\sum\limits_{i} {{{\bf{c}}_i}{{\bf{c}}_i}f_i^{\left( {B1} \right)}}  - \sum\limits_{i} {{{\bf{c}}_i}{{\bf{c}}_i}F_i^{\left( 1 \right)}} } \right]\\&
		=  - {\tau _f}\Delta t\left[ {\frac{\partial }{{\partial {t_1}}}{{\bf{u}}_\alpha }{{\bf{u}}_\beta } + \frac{\partial }{{\partial {t_1}}}p{\delta _{\alpha \beta }} + c_s^2{\nabla _{1\alpha }} \cdot {{\bf{u}}_\beta } + c_s^2{\nabla _{1\beta }} \cdot {{\bf{u}}_\alpha } - \frac{1}{{\Delta t}}\sum\limits_{i} {{{\bf{c}}_i}{{\bf{c}}_i}f_i^{\left( {B1} \right)}}  - \sum\limits_{i} {{{\bf{c}}_i}{{\bf{c}}_i}F_i^{\left( 1 \right)}} } \right]\\&
		=  - {\tau _f}\Delta t\left[ { - {{\bf{u}}_\alpha }{\nabla _{1\beta }} \cdot p{\bf{I}} - {{\bf{u}}_\beta }{\nabla _{1\alpha }} \cdot p{\bf{I}} - {\nabla _{1\gamma }} \cdot \left( {{{\bf{u}}_\alpha }{{\bf{u}}_\beta }{{\bf{u}}_\gamma }} \right) - {{\bf{u}}_\alpha }{{\bf{u}}_\beta }{\nabla _{1\gamma }} \cdot {{\bf{u}}_\gamma } + \frac{\partial }{{\partial {t_1}}}p{\delta _{\alpha \beta }}} \right.\\& \quad \quad \quad \quad
		{\rm{         }}\left. { + {u_\alpha }F_\beta ^{\left( 1 \right)} + {u_\beta }F_\alpha ^{\left( 1 \right)} + c_s^2{\nabla _{1\alpha }} \cdot {{\bf{u}}_\beta } + c_s^2{\nabla _{1\beta }} \cdot {{\bf{u}}_\alpha } - \frac{1}{{\Delta t}}\sum\limits_{i} {{{\bf{c}}_i}{{\bf{c}}_i}f_i^{\left( {B1} \right)}}  - \sum\limits_{i} {{{\bf{c}}_i}{{\bf{c}}_i}F_i^{\left( 1 \right)}} } \right]\\&
		=  - {\tau _f}\Delta t\left[ {{u_\alpha }F_\beta ^{\left( 1 \right)} + {u_\beta }F_\alpha ^{\left( 1 \right)} + c_s^2\left( {{\nabla _{1\alpha }} \cdot {{\bf{u}}_\beta } + {\nabla _{1\beta }} \cdot {{\bf{u}}_\alpha }} \right) - \frac{1}{{\Delta t}}\sum\limits_{i} {{{\bf{c}}_i}{{\bf{c}}_i}f_i^{\left( {B1} \right)}}  - \sum\limits_{i} {{{\bf{c}}_i}{{\bf{c}}_i}F_i^{\left( 1 \right)}} } \right] + O\left( {\Delta tM{a^2}} \right)\\&		
		=  - {\tau _f}\Delta t\left[ {c_s^2\left( {{\nabla _{1\alpha }} \cdot {{\bf{u}}_\beta } + {\nabla _{1\beta }} \cdot {{\bf{u}}_\alpha }} \right) - \frac{1}{{\Delta t}}\sum\limits_{i} {{{\bf{c}}_i}{{\bf{c}}_i}f_i^{\left( {B1} \right)}}  + \frac{1}{{2{\tau _f}}}\left( {{u_\alpha }F_\beta ^{\left( 1 \right)} + {u_\beta }F_\alpha ^{\left( 1 \right)}} \right)} \right] + O\left( {\Delta tM{a^2}} \right).
\end{aligned}
\end{equation}
Under the assumption of low Mach number, the term $O\left( {\Delta tM{a^2}} \right)$ can be neglected. With the aid of the above equation, Eq.(\ref{first_order_epsilon_2}) can be rewritten as
\begin{equation}\label{first_order_epsilon_2_2}
	\frac{\partial }{{\partial {t_2}}}{\bf{u}} = {\tau _f}\Delta t\left( {1 - \frac{1}{{2{\tau _f}}}} \right){\nabla _1} \cdot c_s^2\left( {{\nabla _{1\alpha }} \cdot {{\bf{u}}_\beta } + {\nabla _{1\beta }} \cdot {{\bf{u}}_\alpha }} \right) - {\tau _f}{\nabla _1} \cdot \sum\limits_{i} {{{\bf{c}}_i}{{\bf{c}}_i}f_i^{\left( {B1} \right)}}  - \frac{1}{2}\sum\limits_{i} {{{\bf{c}}_i}f_i^{\left( {B1} \right)}} .
\end{equation}
Summing Eq.(\ref{zero_order_epsilon_1}) on the $t_1$ and $t_2$ scales, one can obtain
\begin{equation}
	\nabla  \cdot {\bf{u}} = \frac{1}{{\Delta t}}\sum\limits_{i} {f_i^B} .
\end{equation}
To recover the mass equation of Eq.(\ref{mass_conservation}), it is obvious that
\begin{equation}\label{zero_order_fB}
	\sum\limits_{i} {f_i^B}  = \sum\limits_{i} {f_i^{\left( {B1} \right)}}  = 0,
\end{equation}
Substituting Eq.(\ref{zero_order_fB}) into Eq.(\ref{zero_order_epsilon_2}), we have
\begin{equation}\label{first_order_fB}
	\sum\limits_{i} {{{\bf{c}}_i}f_i^B}  = \sum\limits_{i} {{{\bf{c}}_i}f_i^{\left( {B1} \right)}}  = 0.
\end{equation}
Taking Eqs.(\ref{first_order_epsilon_1}) and (\ref{first_order_epsilon_2_2}), and combining the equations on the $t_1$ and $t_2$ scales, we obtain
\begin{equation}
	\frac{\partial }{{\partial t}}{\bf{u}} + \nabla  \cdot {\bf{uu}} =  - \nabla  \cdot p{\bf{I}} + \nabla  \cdot \left[ {{\mu _s}\left( {\nabla {\bf{u}} + \nabla {{\bf{u}}^{\rm{T}}}} \right)} \right] - {\tau _f}\nabla  \cdot \sum\limits_{i} {{{\bf{c}}_i}{{\bf{c}}_i}f_i^B}  + {\bf{F}},
\end{equation}
in which
\begin{equation}\label{second_order_fB}
	- {\tau _f}\nabla  \cdot \sum\limits_{i} {{{\bf{c}}_i}{{\bf{c}}_i}f_i^B} = \nabla  \cdot {\tau _p}
\end{equation}
is the divergence of polymer stress tensor. Obviously, Eq.(\ref{second_order_fB}) establishes a connection between the polymer stress tensor in the incompressible momentum conservation equation and the polymer stress fluctuation term ${f_i}^B$. According to the above equations, ${f_i}^B$ must satisfy Eqs.(\ref{zero_order_fB}), (\ref{first_order_fB}) and (\ref{second_order_fB}) to guarantee the conservations of mass and momentum. The identity conditions described above correspond to a system of underdetermined linear equations related to ${f_i}^B$. In this work, we utilize the pseudo-inverse method to determine the polymer stress fluctuation term ${f_i}^B$ (see more details in Appendix A.), which is a widely used computational method for solving systems of underdetermined linear equations \cite{Khoo_IJIE2014,Ba_PRE2018}. By computing the pseudo-inverse matrix, we can obtain the optimal solution that meets the constraints. With the pseudo-inverse method, ${f_i}^B$ are given by (in the case of $c=1$)
\begin{equation}
	\left( {\begin{array}{*{20}{c}}
			{f_0^B}\\
			{f_1^B}\\
			{f_2^B}\\
			{f_3^B}\\
			{f_4^B}\\
			{f_5^B}\\
			{f_6^B}\\
			{f_7^B}\\
			{f_8^B}
	\end{array}} \right) =  - \frac{1}{{72 {\tau _f}}}\left( {\begin{array}{*{20}{c}}
			0&{ - 24}&0&{ - 24}\\
			0&{12}&0&{ - 24}\\
			0&{ - 24}&0&{12}\\
			0&{12}&0&{ - 24}\\
			0&{ - 24}&0&{12}\\
			9&{12}&9&{12}\\
			{ - 9}&{12}&{ - 9}&{12}\\
			9&{12}&9&{12}\\
			{ - 9}&{12}&{ - 9}&{12}
	\end{array}} \right)\left( {\begin{array}{*{20}{c}}
			{{\tau _{{p_{yx}}}}}\\
			{{\tau _{{p_{xx}}}}}\\
			{{\tau _{{p_{xy}}}}}\\
			{{\tau _{{p_{yy}}}}}
	\end{array}} \right),
\end{equation}
in which ${\tau _{{p_{ij}}}}\left( {i,j = x,y} \right)$ is the component of the polymer stress tensor. In such a scenario, the influence of polymer stress is reflected in the ${f_i}^B$, thereby avoiding the computation of the divergence of the stress tensor. In the more complex three-dimensional case, we present the computation of ${f_i}^B$ in Appendix A.

\subsection{Lattice Boltzmann model for the Oldroyd-B constitutive equation}

To simulate the evolution of the stress tensor in Eq.(\ref{constitutive_equation}), we calculate each component of the conformation tensor using an advection-diffusion scheme that substitutes the scalar density distribution ${f_i}({\bf{x}},t)$ with a tensor distribution ${h_{\alpha \beta i}}({\bf{x}},t)$ \cite{Shi_PRE2009}. The evolution equation of the stress field distribution function is given by \cite{Su_JNNFM2013}
\begin{equation}
	{h_{\alpha \beta i}}\left( {{\bf{x}} + {{\bf{c}}_i}\Delta t,t + \Delta t} \right) - {h_{\alpha \beta i}}({\bf{x}},t) =  - \frac{1}{{{\tau _h}}}\left[ {{h_{\alpha \beta i}}({\bf{x}},t) - h_{\alpha \beta i}^{({\rm{eq}})}({\bf{x}},t)} \right] + \Delta t{\chi _{\alpha \beta i}}\left( {{\bf{x}},t} \right) + 0.5{\left( {\Delta t} \right)^2}\frac{{\partial {\chi _{\alpha \beta i}}}}{{\partial t}},
\end{equation}
in which ${h_{\alpha \beta i}}$ denotes the distribution function of the component ${{{{A}}}_{\alpha \beta }}$ of the conformation tensor ${\bf{{A}}}$ along the $i{\rm{ - th}}$ discrete particle velocity; ${{\tau _h}}$ and $\Delta t$ are the relaxation parameter for the stress evolution and the streaming time step, respectively. $h_{\alpha \beta i}^{({\rm{eq}})}$ represents the equilibrium distribution function which is defined as
\begin{equation}
	h_{\alpha \beta i}^{({\rm{eq}})} = {\omega _i}{{{A}}_{\alpha \beta }}\left[ {1 + \frac{{{\bf{c}}_i \cdot {\bf{u}}}}{{c_s^2}}} \right].
\end{equation}
The source term in this LB scheme related to the constitutive equation is given by
\begin{equation}
	{\chi _{\alpha \beta i}} = {\omega _i}{\chi _{\alpha \beta }}\left[ {1 + \left( {1 - \frac{1}{{2{\tau _h}}}} \right)\frac{{{{\bf{c}}_i} \cdot {\bf{u}}}}{{c_s^2}}} \right],
\end{equation}
${\chi _{\alpha \beta }}$ is the component of ${\boldsymbol{\chi }}$, which is defined as
\begin{equation}
	{\boldsymbol{\chi }} =  - \frac{1}{\lambda }\left( {{\bf{A}} - {\bf{I}}} \right) + {\bf{A}} \cdot \nabla {\bf{u}} + {\left( {\nabla {\bf{u}}} \right)^{\rm{T}}} \cdot {\bf{A}}.
\end{equation}
${A_{\alpha \beta }}$ is computed by
\begin{equation}
	{A_{\alpha \beta }} = \sum\limits_i {{h_{\alpha \beta i}}} .
\end{equation}

\section{Numerical Examples}\label{section_3}

In this section, several benchmark examples including the simplified four-roll mill, planar poiseuille flow, the unsteady Womersley flow, and the three-dimensional Taylor–Green vortex are performed to validate the reliability of the present model. In these cases, fluid flow behavior is characterized by three key parameters: the kinematic viscosity ratio $\beta$, the Weissenberg number $Wi$, and the Reynolds number $Re$. These parameters are crucial for describing the flow behavior of viscoelastic fluid and are defined as
\begin{equation}
	\beta  = \frac{{{\nu _s}}}{{{\nu _s} + {\nu _p}}},\quad  {\rm{   }}Wi = \lambda \dot \gamma  = \frac{{\lambda U}}{L},\quad {\rm{   }}{\mathop{\rm Re}\nolimits}  = \frac{{UL}}{{{\nu _s} + {\nu _p}}},
\end{equation}
in which $\nu_s$ and $\nu_p$ are the kinematic viscosity of the solvent and polymer, respectively. $\dot \gamma$ denotes the characteristic shear rate, and $U$ and $L$ represent the characteristic velocity and length of the flow, respectively. In our simulations, the relaxtion parameter is fixed at $\tau_h=0.501$, and all tests are carried out on the Intel Xeon Silver 4512R equipped with 128 GB of CPU memory.

\subsection{Simplified four-roll mill}

\begin{figure}[H]
	\centering
	\includegraphics[width=0.36\textwidth]{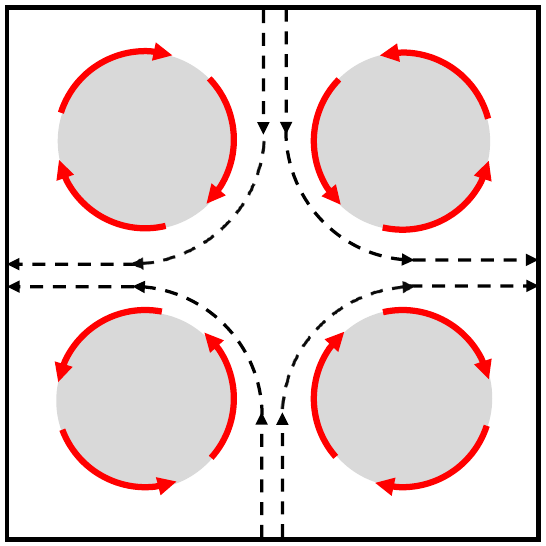}
	\caption{Schematic of simplified four-roll mill.}
	\label{Four_rolling_mill}
\end{figure}

This test case simulates the effect of creating an elongational flow near the center stagnation point between four rotating cylinders. The physical domain is set to be $\left[ {0,2\pi } \right] \times \left[ {0,2\pi } \right]$, and the cylinders rotate in the direction indicated by the red arrow, as shown in Fig. \ref{Four_rolling_mill}. We utilize a body force to simplify the effects of the rotating cylinders, and the body force $\bf{g}$ is defined as 
\begin{equation}
	{g_x} = 2{\nu _s}{U_0}\sin \left( x \right)\cos \left( y \right),
\end{equation}
\begin{equation}
	{g_y} =  - 2{\nu _s}{U_0}\cos \left( x \right)\sin \left( y \right),
\end{equation}
the velocity field of the corresponding Newtonian fluid flow at low Reynolds number is described by
\begin{equation}\label{velocity_init}
	\mathbf{U}=U_0(\sin (x) \cos (y),-\cos (x) \sin (y)),
\end{equation}
in which $U_0$ represents the maximum velocity in the whole domain and is regarded as the characteristic velocity. The velocity field near the center stagnation point $(\pi,\pi)$ is given by
\begin{equation}
	{\bf{U}}\left( {\pi ,\pi } \right) = {U_0}\left( {\epsilon , - \epsilon } \right) = \left( {\frac{{\partial {u_x}\left( {\pi ,\pi } \right)}}{{\partial x}},\frac{{\partial {u_y}\left( {\pi ,\pi } \right)}}{{\partial y}}} \right),
\end{equation}
where $\epsilon$ denotes the dimensionless local elongational rate. By multiplying the Weissenberg number with the local elongational rate, we calculate the effective Weissenberg number, which is defined as
\begin{equation}
	W{i^{effect}} = \epsilon Wi.
\end{equation}
Thomases and Shelley et al. demonstrated that when the viscoelastic flow reaches a steady state, the components of the conformation tensor $\bf{A}$ can be approximated by the following solution
\begin{subequations}
	\begin{align}
		&{A_{xx}}\left( {\pi ,y} \right) = \frac{1}{{1 - 2W{i^{effect}}}} + C{\left| y \right|^{\frac{{\left( {1 - 2W{i^{effect}}} \right)}}{{W{i^{effect}}}}}},\\
		&{A_{yy}}\left( {\pi ,y} \right) = \frac{1}{{1 + 2W{i^{effect}}}} + {\left| y \right|^{\left( {2 + \frac{1}{{W{i^{effect}}}}} \right)}},\\
		&{A_{xy}}\left( {\pi ,y} \right) = 0,
	\end{align}
\end{subequations}
in which $C$ represents a constant term that requires adjustment. In our simulations, the whole computational domain has a size of $200lu \times 200lu$. The initialization of the velocity field is determined by Eq.(\ref{velocity_init}), and the polymer stress tensor is initially set to zero. Since the present simulations emphasize viscoelastic behavior at low Reynolds numbers, the Reynolds number is set to $Re=1.0$. Additionally, the kinematic viscosity ratio is fixed at $\beta=2/3$ to align with previous studies. The characteristic length is defined as $L=1.0$, and the characteristic velocity is set to $U_0$ to ensure that the numerical Mach number $Ma = {U}/c_s^2 \le 0.15$. As for the boundary conditions, we employ the periodic boundary scheme for both the horizontal and vertical directions.

\begin{figure}[H]
	\centering
	\includegraphics[scale=0.36]{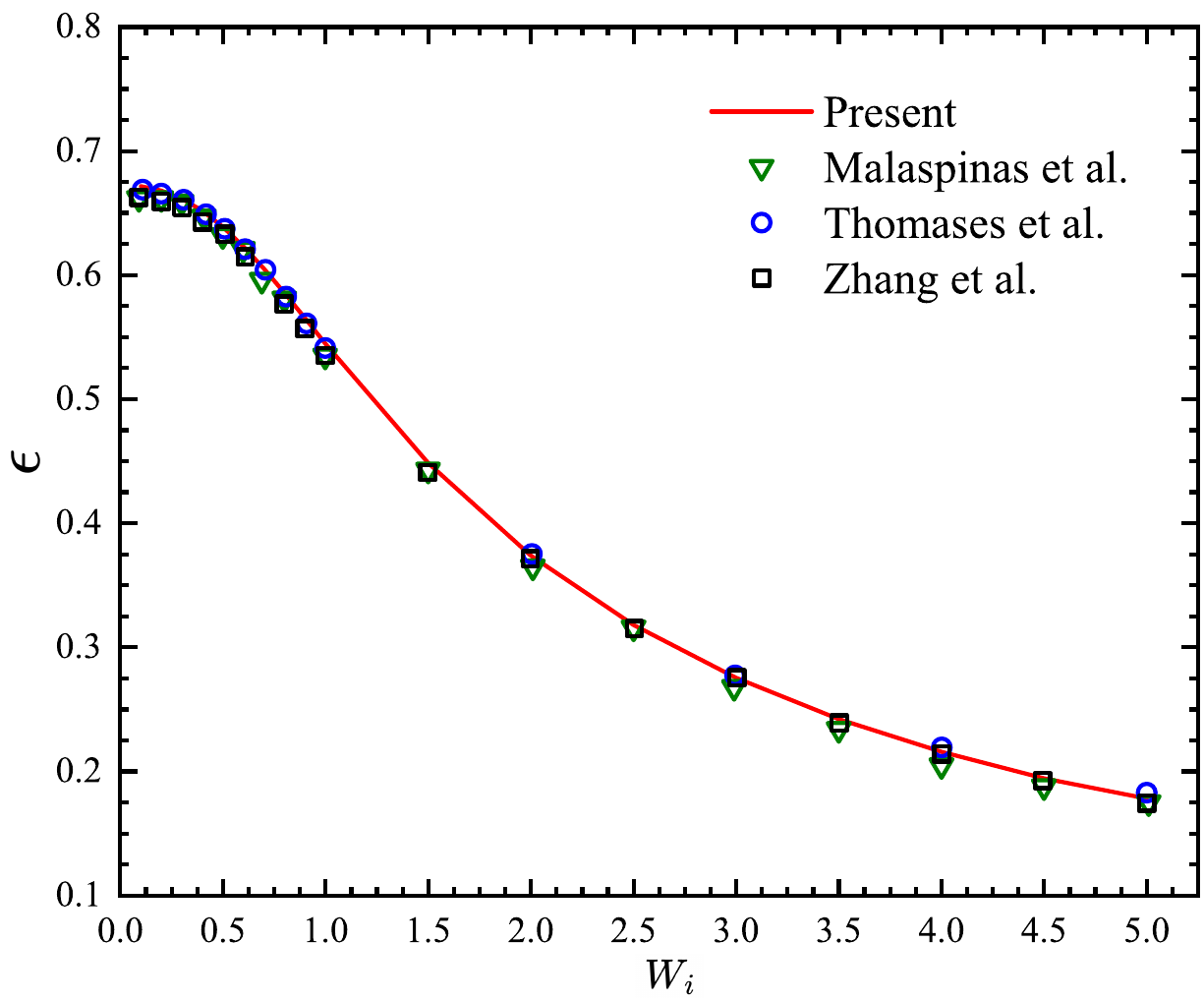} 
	\put(-220 ,170){(\textit{a})}
	\quad
	\includegraphics[scale=0.36]{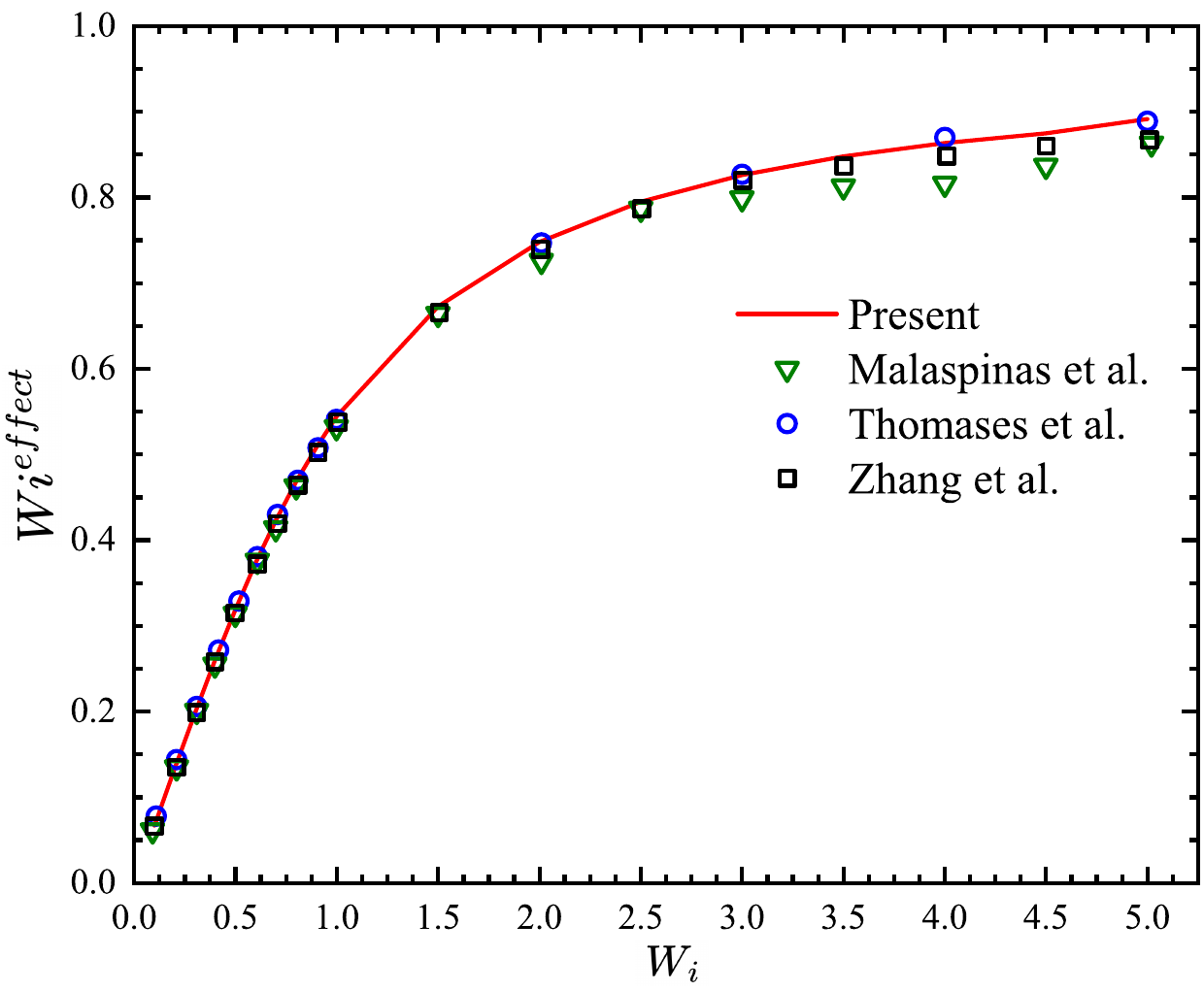} 
	\put(-220 ,170){(\textit{b})}
	\caption{The comparison of (a) the local elongational rate and (b) effective $Wi$ for various $Wi$.}
	\label{Elongational_and_Wi_effect}
\end{figure}

\begin{figure}[H]
	\centering
	\includegraphics[width=0.46\textwidth]{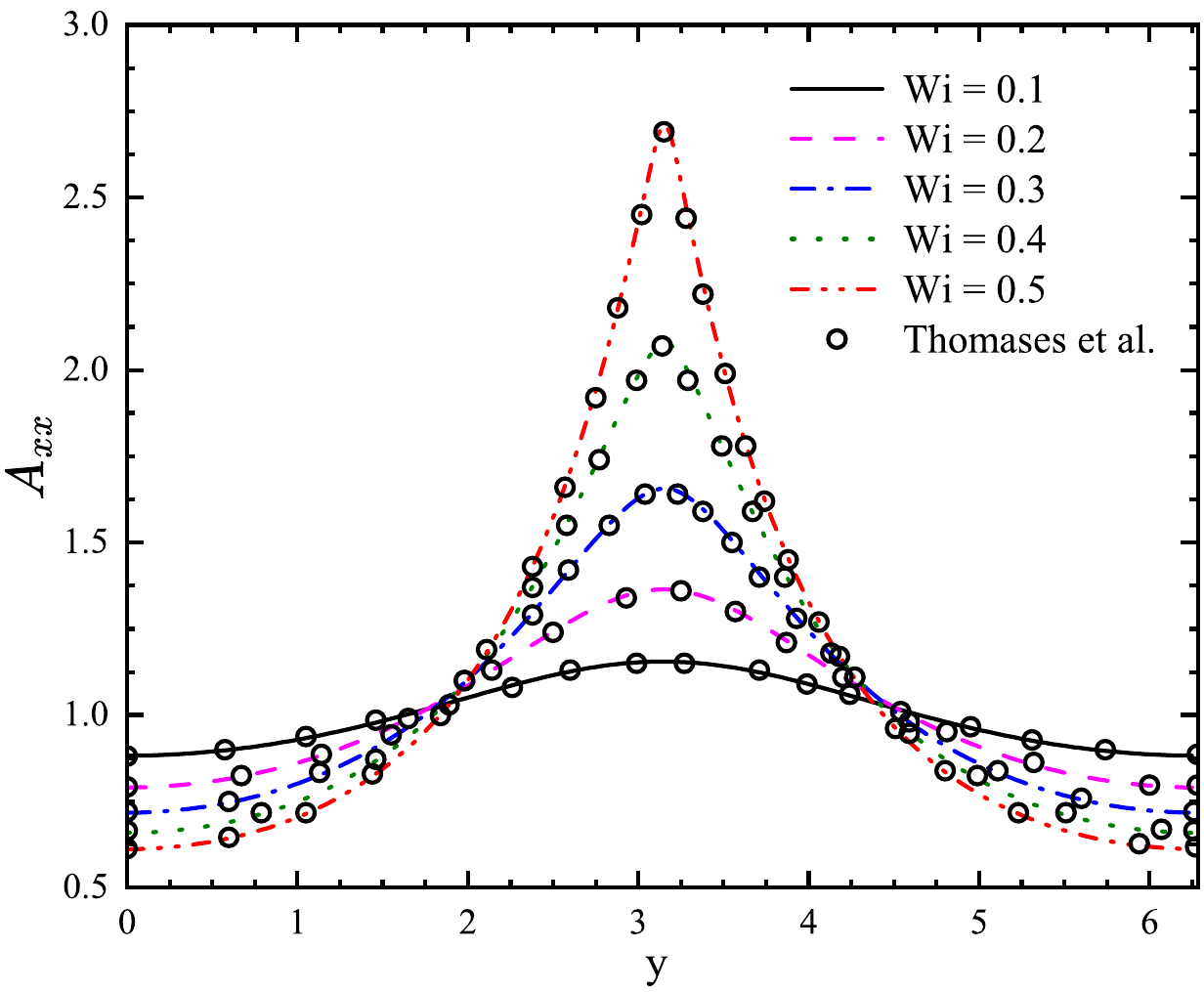}
	\caption{The variation of the conformation tensor $A_{xx}$ along the vertical centerline for different $Wi$.}
	\label{Axx_comparison}	
\end{figure}

\begin{figure}[H]
	\centering
	\begin{flushleft}
		\includegraphics[scale=0.3]{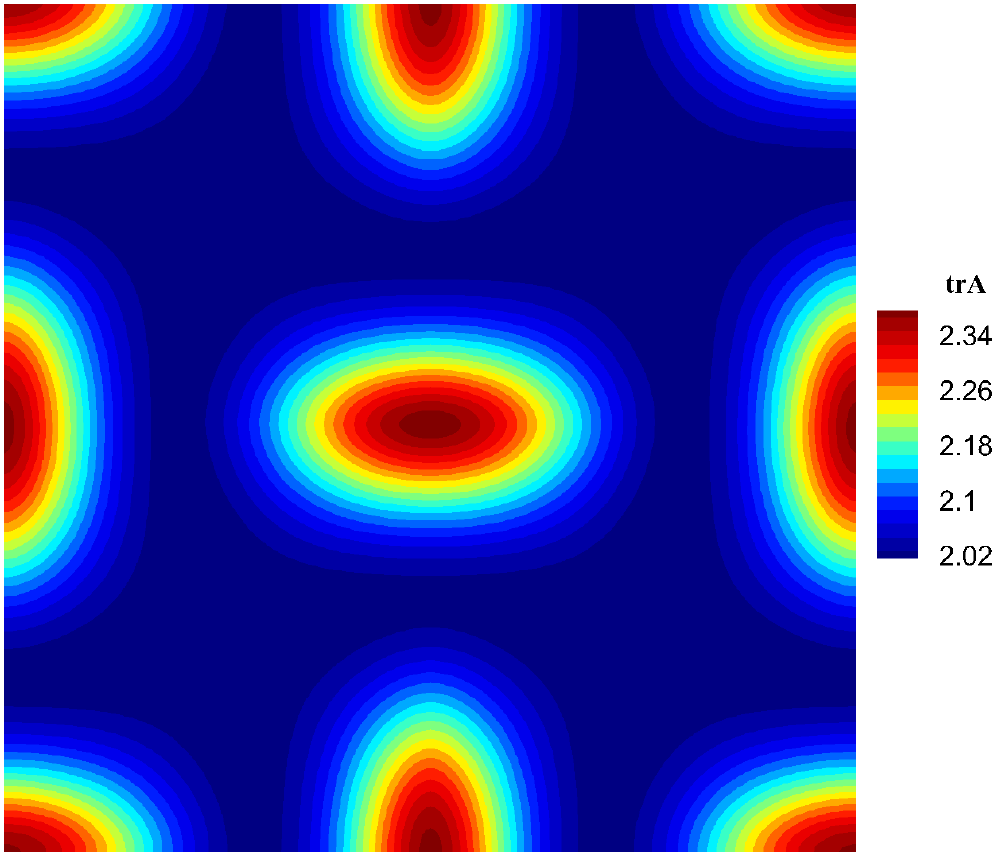}
		\put(-160 ,115){(\textit{a})}
		\put(-185, 55){\textit{$Wi=0.3$}} \quad
		\includegraphics[scale=0.3]{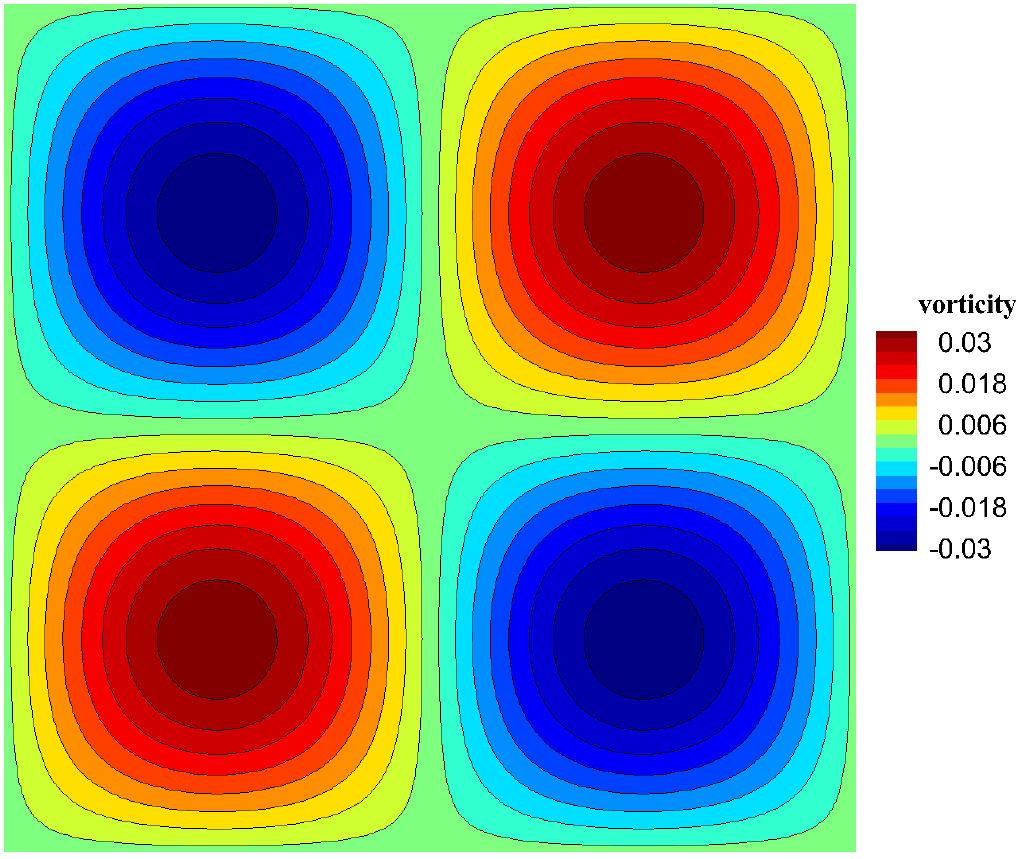}
		\put(-160 ,115){(\textit{b})} \quad
		\includegraphics[scale=0.3]{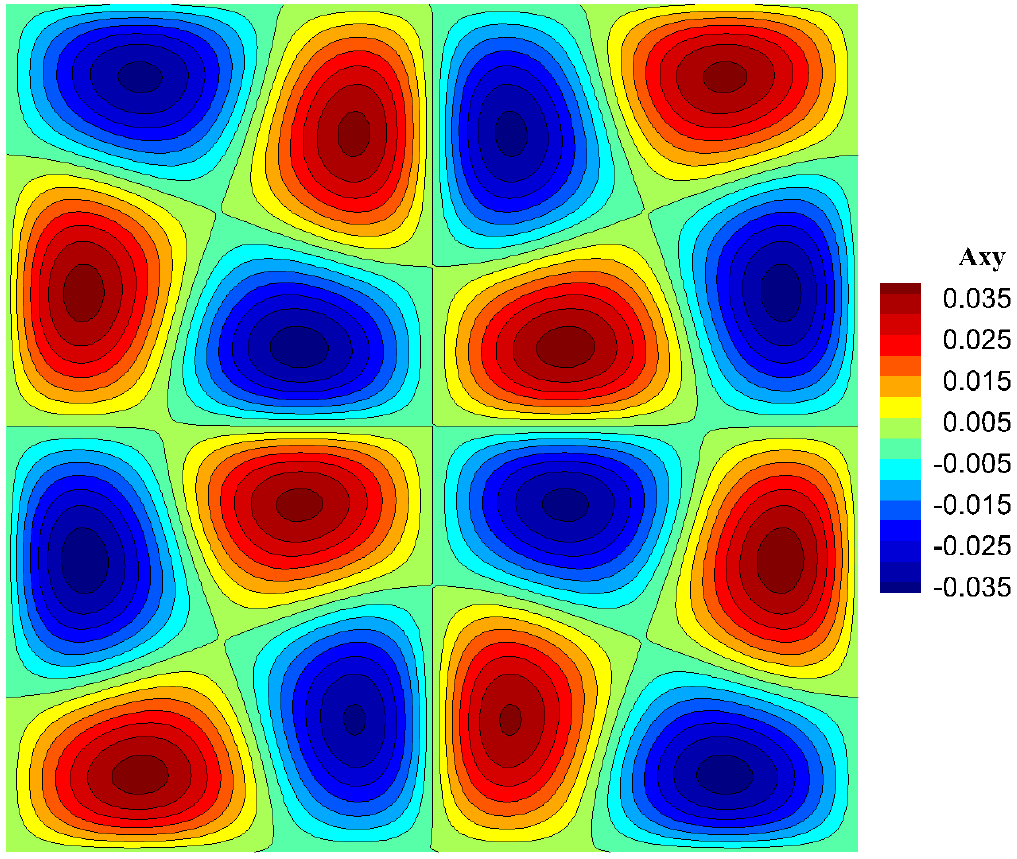}
		\put(-160 ,115){(\textit{c})} \\
		
		\includegraphics[scale=0.3]{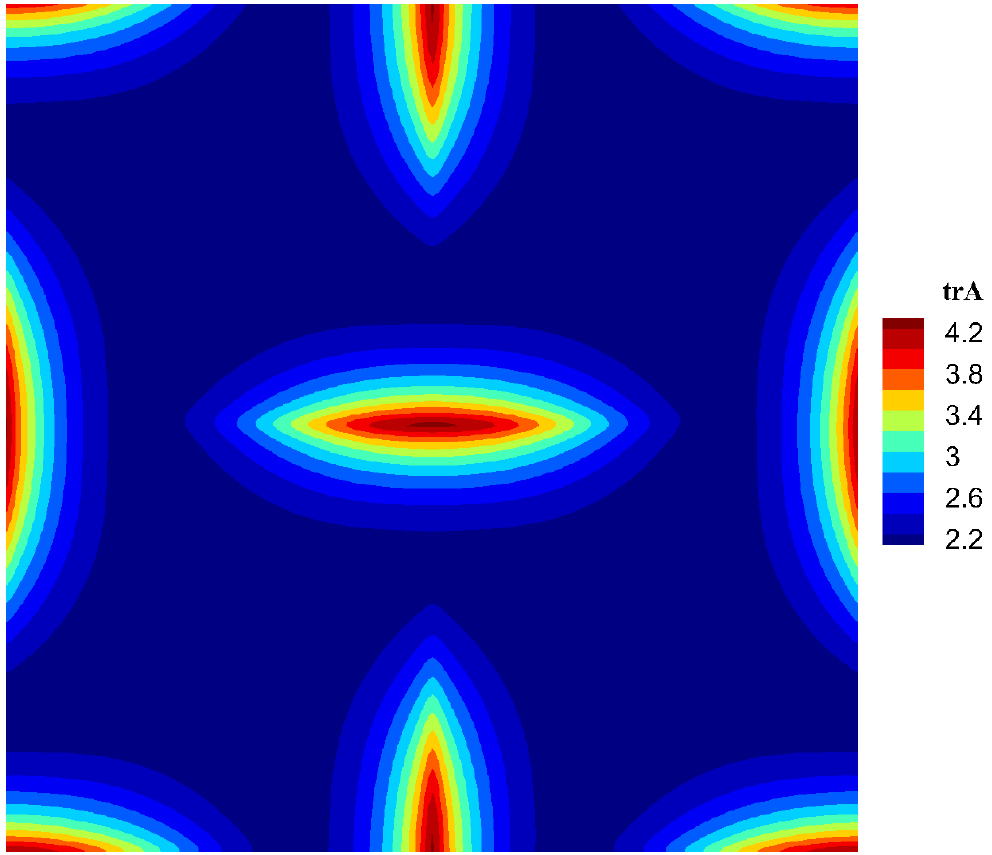}
		\put(-160 ,115){(\textit{d})}
		\put(-185, 55){\textit{$Wi=0.6$}} \quad
		\includegraphics[scale=0.3]{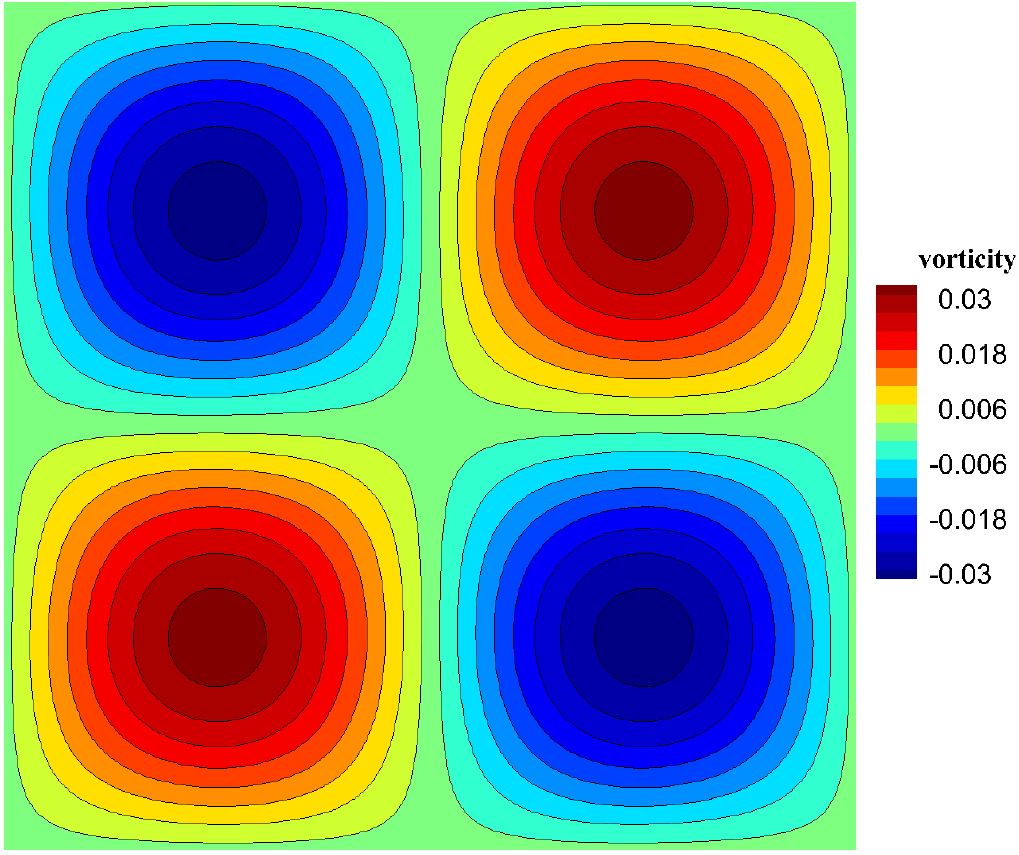}
		\put(-160 ,115){(\textit{e})} \quad
		\includegraphics[scale=0.3]{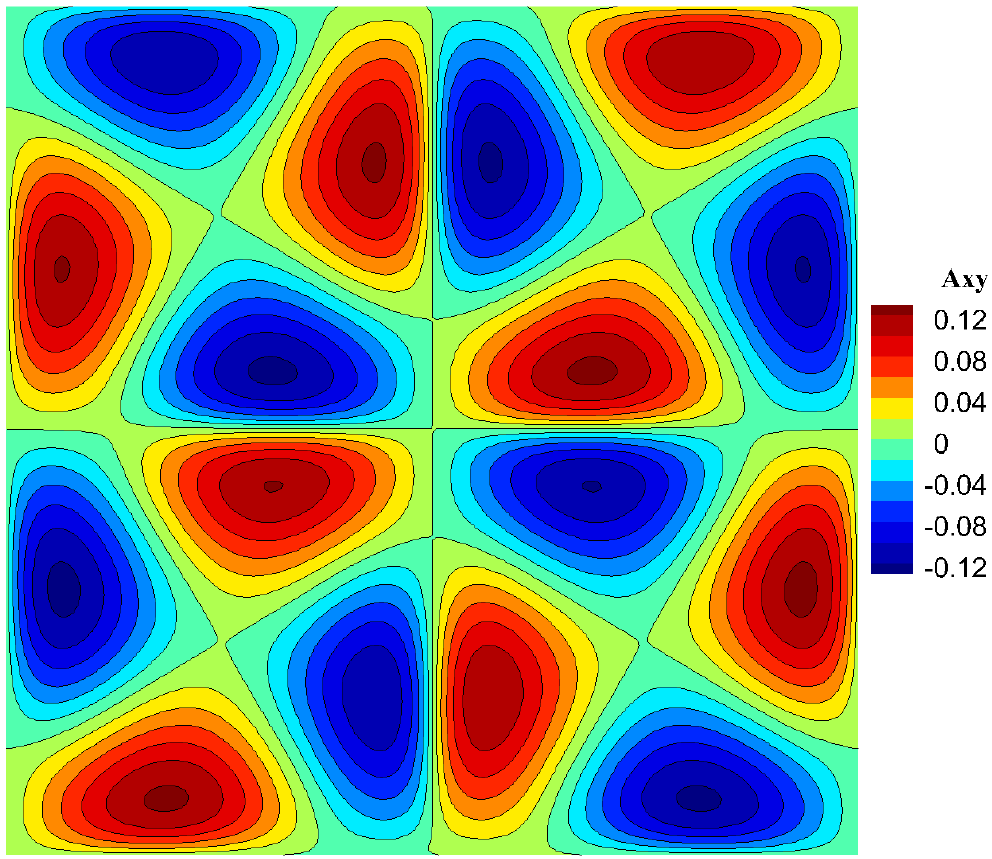} 
		\put(-155 ,115){(\textit{f})} \\
		
		\includegraphics[scale=0.3]{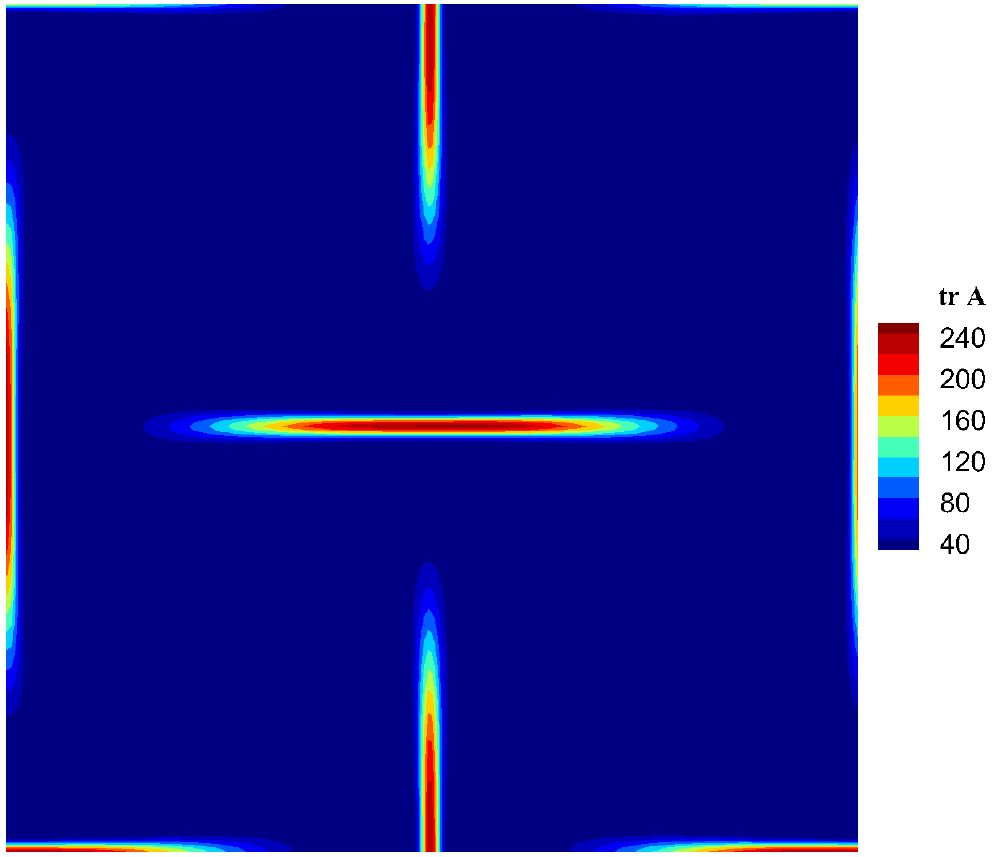}
		\put(-160 ,115){(\textit{g})}
		\put(-185, 55){\textit{$Wi=5.0$}} \quad
		\includegraphics[scale=0.3]{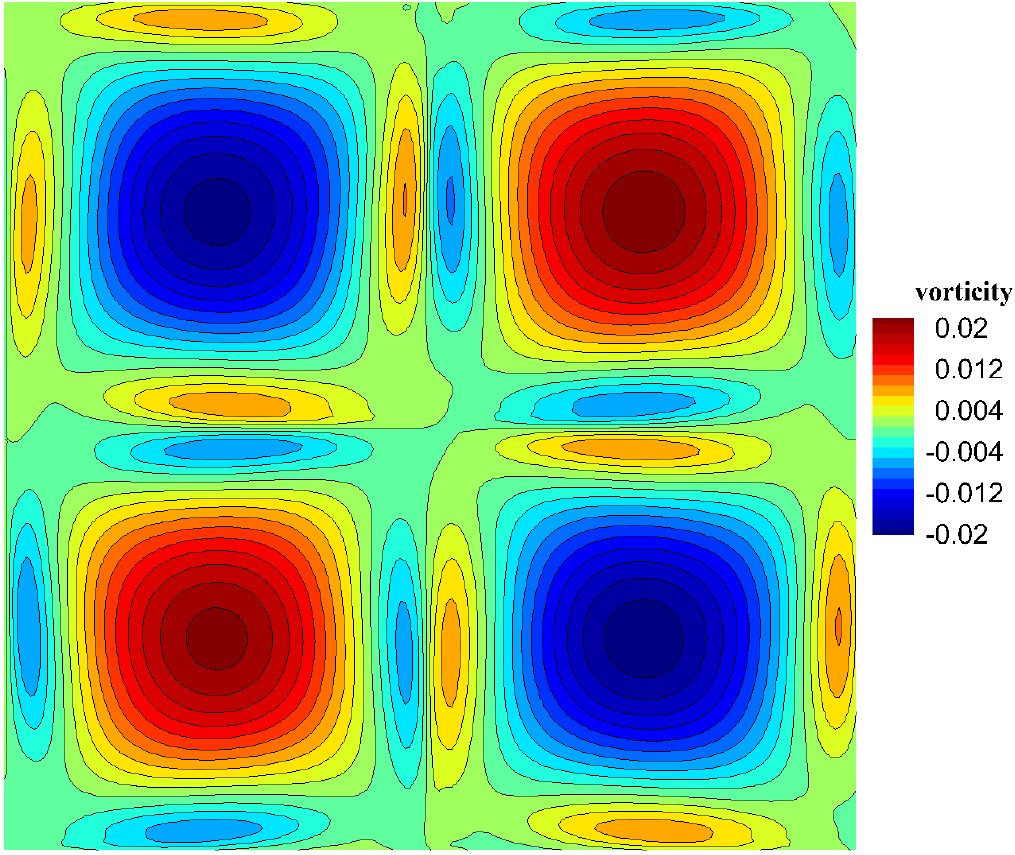}
		\put(-160 ,115){(\textit{h})} \quad
		\includegraphics[scale=0.3]{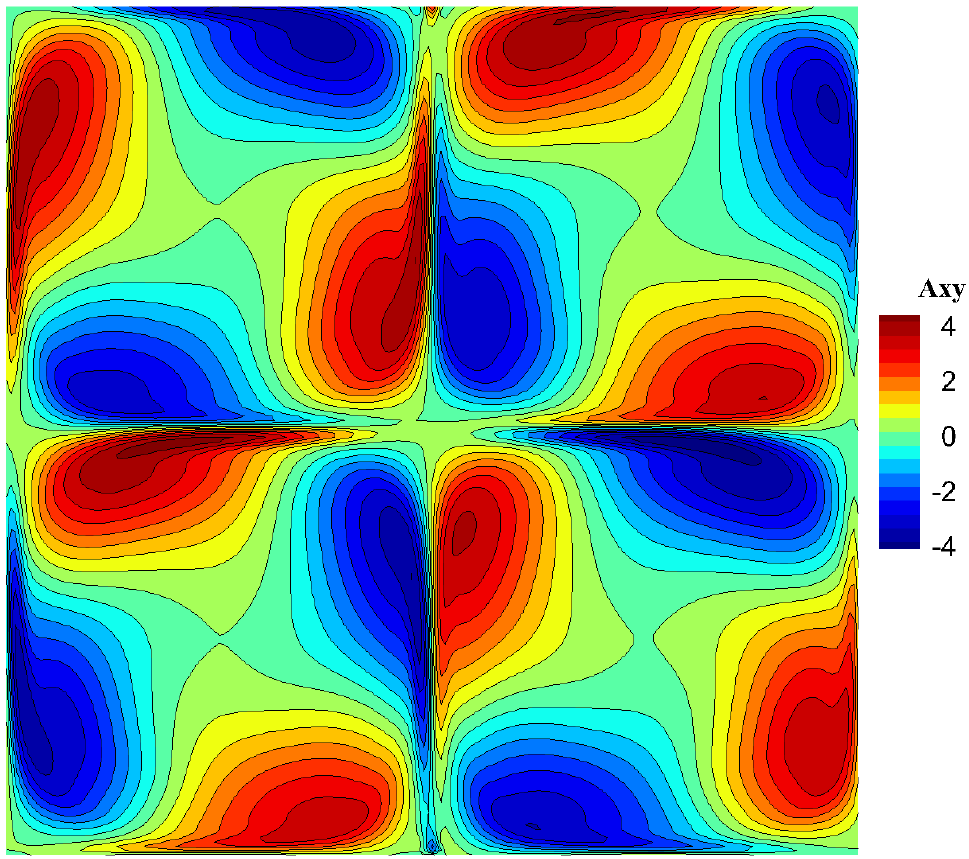}
		\put(-153 ,115){(\textit{i})} \\
	\end{flushleft}	
	\caption{Contour plots of ${\rm{tr}\bf{A}}$, vorticity field and $A_{xy}$ (left to right) for $Wi=0.3$ (a-c), $Wi=0.6$ (d-f), and $Wi=5.0$ (g-i).}
	\label{trA_vorticity_Axy}
\end{figure}

To validate the capability of the current model in simulating viscoelastic fluid, we calculate the local elongational rate and the effective Weissenberg number for various Weissenberg numbers. We compare the numerical results with those obtained from a pseudospectral method employed by Thomases et al. \cite{Thomases_POF2007}, the LB method used by Malaspinas et al. \cite{Malaspinas_JNNFM2010}, and a viscoelastic lattice Boltzmann flux solver applied by Zhang et al. \cite{Zhang_CF2025}. Fig. \ref{Elongational_and_Wi_effect} (a) and (b) show the changes in the local elongational rate $\epsilon$ and the effective Weissenberg number $W{i^{effect}}$ as the Weissenberg number $Wi$ varies, respectively. It is observed that the local elongational rate $\epsilon$ decreases as the Weissenberg number $Wi$ increases. The results of the present model are in good agreement with those obtained from the previous approaches, which confirm that the current model correctively maintains the hydrodynamic field, and we also evaluate its capability to retain the stress field. Fig. \ref{Axx_comparison} displays the variation of the conformation tensor $A_{xx}$ along the vertical centerline for Weissenberg number $Wi$ ranging from 0.1 to 0.5, and the results obtained from the current model align closely with those reported by Thomases et al. \cite{Thomases_POF2007}.
Fig. \ref{trA_vorticity_Axy} illustrates the contour plots of ${\rm{tr}\bf{A}}=A_{xx}+A_{yy}$, vorticity, and $A_{xy}$ for Weissenberg numbers of 0.3, 0.6, and 5.0. As shown in Fig. \ref{trA_vorticity_Axy}(a)-(c), ${\rm{tr}\bf{A}}$ exhibits a bounded steady solution for $Wi = 0.3$, and the vorticity field retains the characteristics of the four-vortex structure induced by the body force. When $Wi$ increases to 0.6 (see Fig. \ref{trA_vorticity_Axy}(d)-(f)), the value of ${\rm{tr}\bf{A}}$ becomes more concentrated along the extensional direction. Notably, the vorticity field does not change significantly because the velocity field remains predominantly influenced by the body force rather than polymer stress. At $Wi = 5.0$ (see Fig. \ref{trA_vorticity_Axy}(g)-(i)), the polymer stress increases rapidly and concentrates along the extensional direction. Furthermore, Fig. \ref{trA_vorticity_Axy}(h) shows that smaller oppositely signed vortices appear between the rolls.

\subsection{Planar Poiseuille flow}
\begin{figure}[H]
	\centering
	\includegraphics[width=0.46\textwidth]{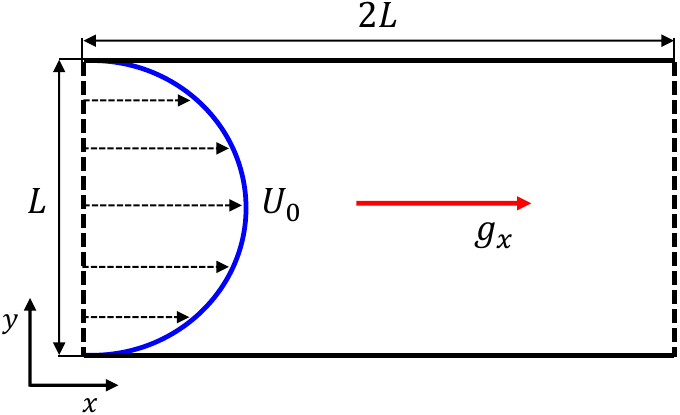}
	\caption{Schematic of Poiseuille flow.}
	\label{Poiseuille_flow}
\end{figure}
\begin{figure}[H]
	\centering
	\includegraphics[width=0.46\textwidth]{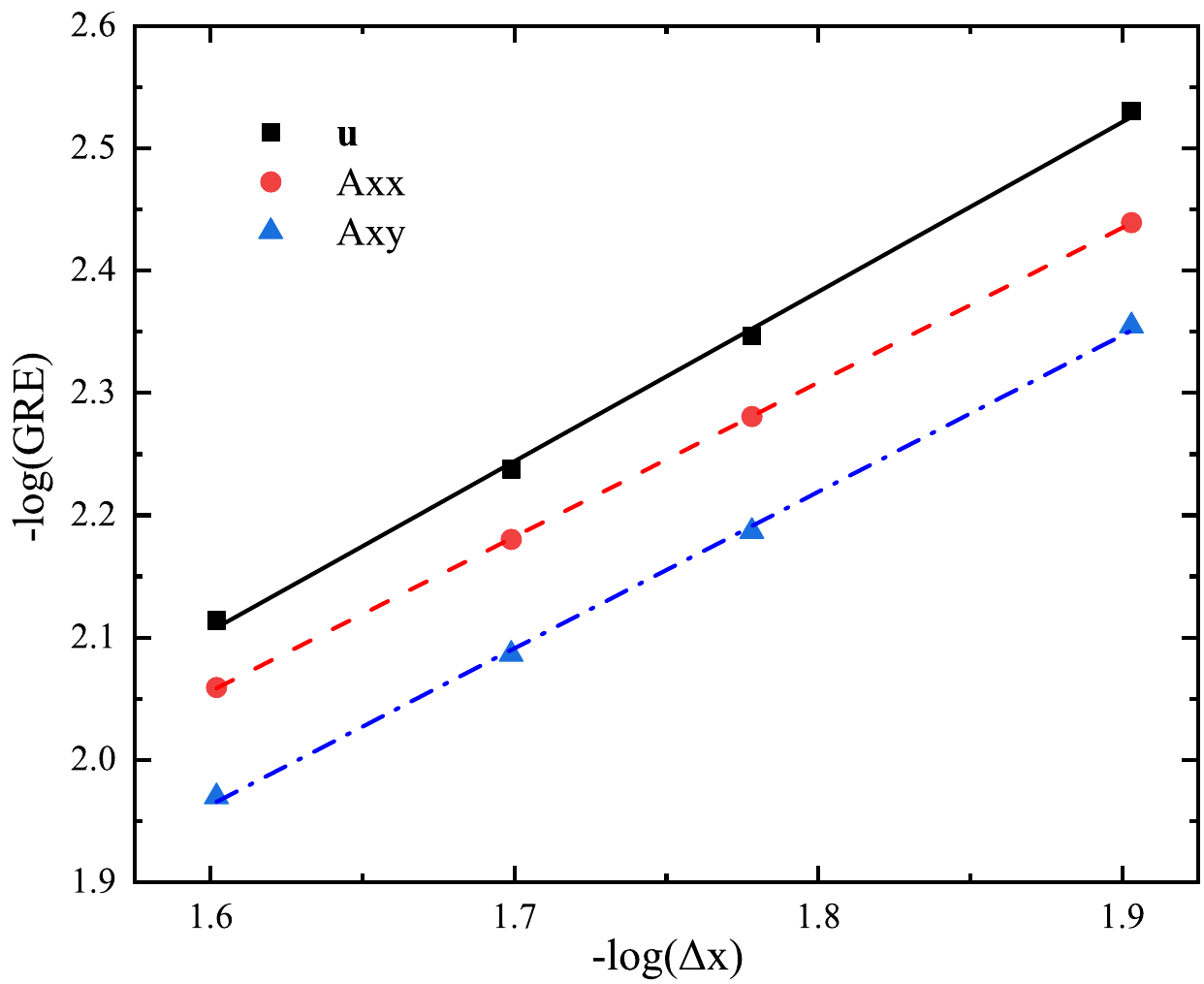}
	\caption{GRE of the present model at various lattice sizes.}
	\label{Poiseuille_flow_error}
\end{figure}

The planar Poiseuille flow is a fundamental benchmark for validating numerical methods, particularly for studying viscoelastic fluid. This flow configuration is well-suited for verification purposes, as analytical solutions exist for certain simplified constitutive equations, such as the Oldroyd-B equation, which provides a reliable basis for comparing numerical results. As shown in Fig. \ref{Poiseuille_flow}, the geometry of the computational domain features a channel with a constant body force $g_x$ imposed in the x-direction to drive the flow. In our simulations, the computational domain measures $256lu \times 128lu$. The periodic boundary condition is implemented at both the inlet and outlet of the channel, while the no-slip boundary condition is applied to the upper and lower walls \cite{Su_PRE2013}. The Weissenberg and Reynolds numbers are defined as $Wi=1.0$, $Re=1.0$, with the characteristic length $L$ set to 1.0, and the characteristic velocity $U$ corresponding to the maximum steady centerline flow velocity $U_0$ at $y = L/2$. Initially, both the velocity and stress tensor of the fluid are set to zero. Driven by the constant body force $g_x$ in the x-direction, the fluid in the channel accelerates from a stationary state and reaches a steady shear flow due to viscous and viscoelastic effects eventually. The analytical solution of the Oldroyd-B constitutive equation in the steady state can be written as
\begin{subequations}
	\begin{gather}
		 u_x=-\frac{g_x}{2}\left(\frac{y^2-y L}{v_s+v_p}\right), \quad u_y=0, \\
		 A_{x x}=1+\frac{g_x^2 \lambda^2 v_p\left(-2 y+L\right)^2}{2\left(v_s+v_p\right)^2}, \\
		 A_{x y}=-\frac{g_x}{2} \frac{\lambda\left(2 y-L\right)}{v_s+v_p}, \\
		 A_{y y}=1 .
	\end{gather}
\end{subequations}
The constant body force $g_x$ is given by
\begin{equation}
	{g_x} = \frac{8}{{{L^2}}}\left( {{\nu _s} + {\nu _p}} \right){U_0}.
\end{equation}
The exact transient solutions for the velocity and viscoelastic stress tensor of the Oldroyd-B model can be found in works \cite{Waters_1970,Carew_1994}, where the transient exact solution for the velocity can be expressed as \cite{Zhang_CF2025}
\begin{equation}\label{poi_transient_solution}
	u_x^*\left( {{y^*},{t^*}} \right) = 4{y^*}\left( {1 - {y^*}} \right) - 32\sum\limits_{n = 1}^{ + \infty } {\frac{{\sin \left( {N{y^*}} \right)}}{{{N^3}}}} {G_N}\left( {El,{t^*}} \right),
\end{equation}
in which
\begin{equation}
	{G_N}\left( {El,{t^*}} \right) = \left\{ {\begin{array}{*{20}{c}}
			{\frac{1}{2}\left( {{a_N}\exp \left( {{p_N}{t^*}} \right) + {b_N}\exp \left( {{q_N}t} \right)} \right)}&{\beta _N^2 \ge 0,}\\
			{\exp \left( { - \alpha _N^*{t^*}} \right)\left( {\cos \left( {\beta _N^*{t^*}} \right) + \frac{{{S_N}}}{{{\beta _N}}}\sin \left( {\beta _N^*{t^*}} \right)} \right)}&{\beta _N^2 < 0,}
	\end{array}} \right.
\end{equation}
with
\begin{subequations}
	\begin{gather}
		 E l=W i / R e, N=(2 n-1) \pi, S_N=1-(2-\beta) N^2 E l, \\
		 \alpha_N=1+\beta E_l N^2, \beta_N^2=\alpha_N^2-4 N^2 E l, \beta_N=\sqrt{\beta_N^{2}}, \\
		 \alpha_N^*=\alpha_N /(2 E l), \beta_N^*=\beta_N /(2 E l), \\
		 a_N=1+S_N / \beta_N, b_N=1-S_N / \beta_N, \\
		 p_N=-\alpha_N^*+\beta_N^*, q_N=-\alpha_N^*-\beta_N^* .
	\end{gather}
\end{subequations}
To verify the spatial accuracy of the current model, we calculate the global relative error (GRE) of $\bf{u}$, $A_{xy}$, and $A_{xx}$ with various grid resolutions. The GRE of $\bf{u}$ is defined as \cite{Chai_JSC2016}
\begin{equation}
	{\rm{GRE}}\left( {{\bf{u}},t} \right) = \frac{{\sum\limits_{i = 1}^M {\left| {{{\bf{u}}_{numerical}}\left( {{x_i},t} \right) - {{\bf{u}}_{analytical}}\left( {{x_i},t} \right)} \right|} }}{{\sum\limits_{i = 1}^M {\left| {{{\bf{u}}_{analytical}}\left( {{x_i},t} \right)} \right|} }},
\end{equation}
in which $i$ represents the index of position and $M$ denotes the total number of the computational nodes. We set $\beta = 0.1$ in our simulations. Fig. \ref{Poiseuille_flow_error} illustrates the GRE for both the velocity and stress fields at a steady state. The slopes of the fitting lines for the various results are nearly equal to 2.0, indicating that the present model achieves a second-order convergence rate in space.

\begin{figure}[H]
	\centering
	\includegraphics[width=0.46\textwidth]{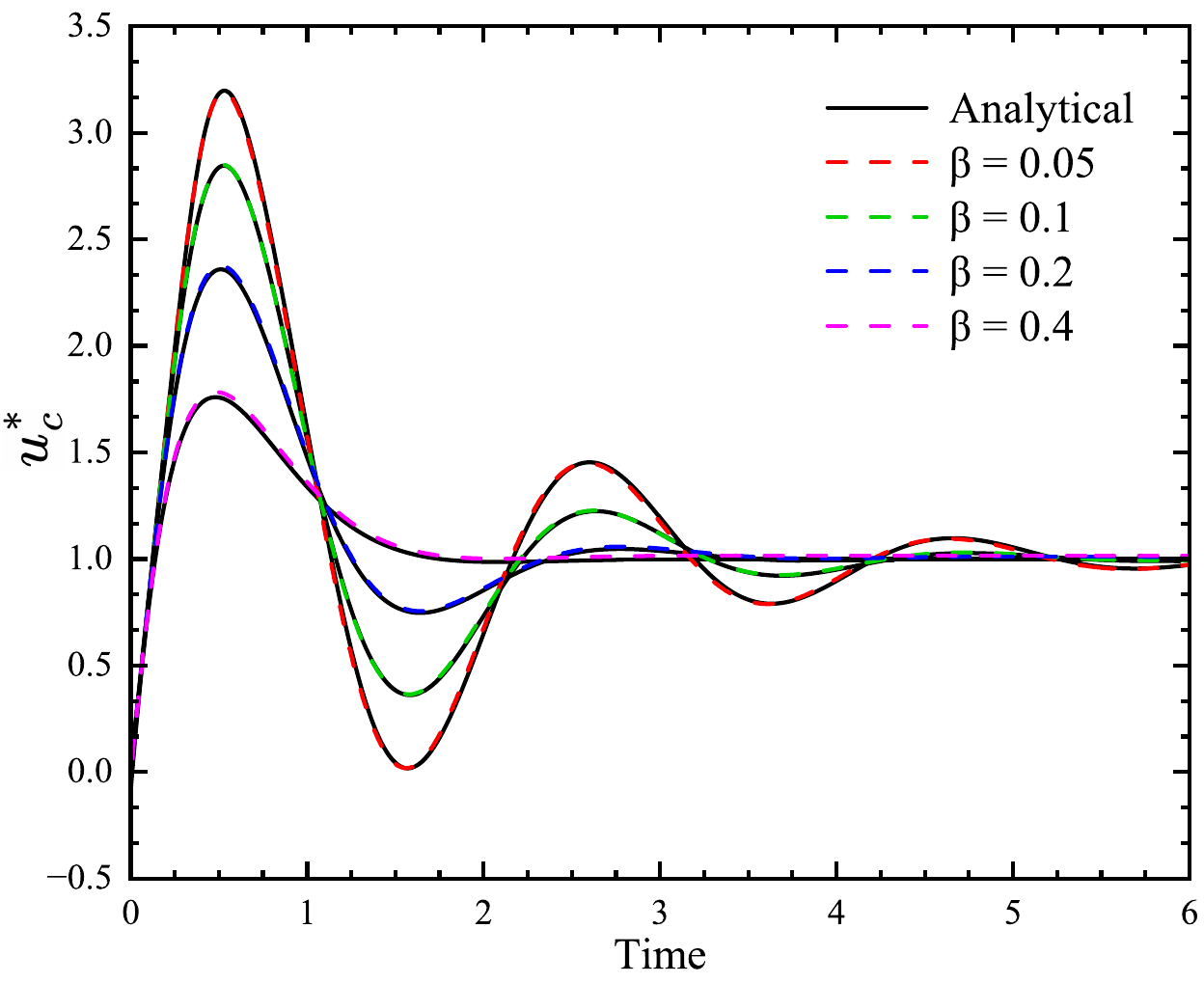}
	\caption{The comparison of analytical and numerical transient centerline velocities for various kinematic viscosity ratios.}
	\label{beta_Uc_transient}
\end{figure}

\begin{figure}[H]
	\centering
	\includegraphics[scale=0.36]{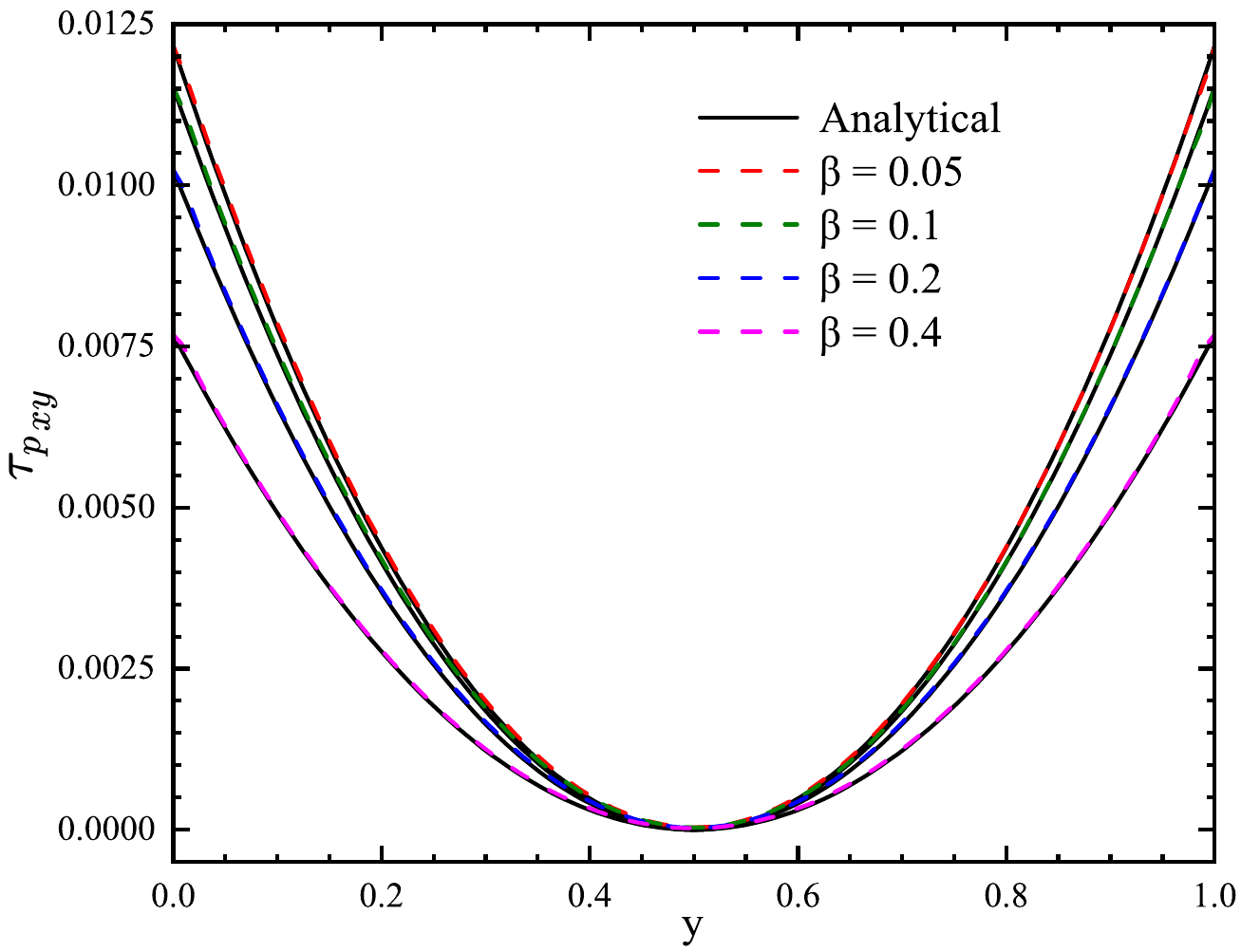} 
	\put(-230 ,168){(\textit{a})}
	\quad
	\includegraphics[scale=0.36]{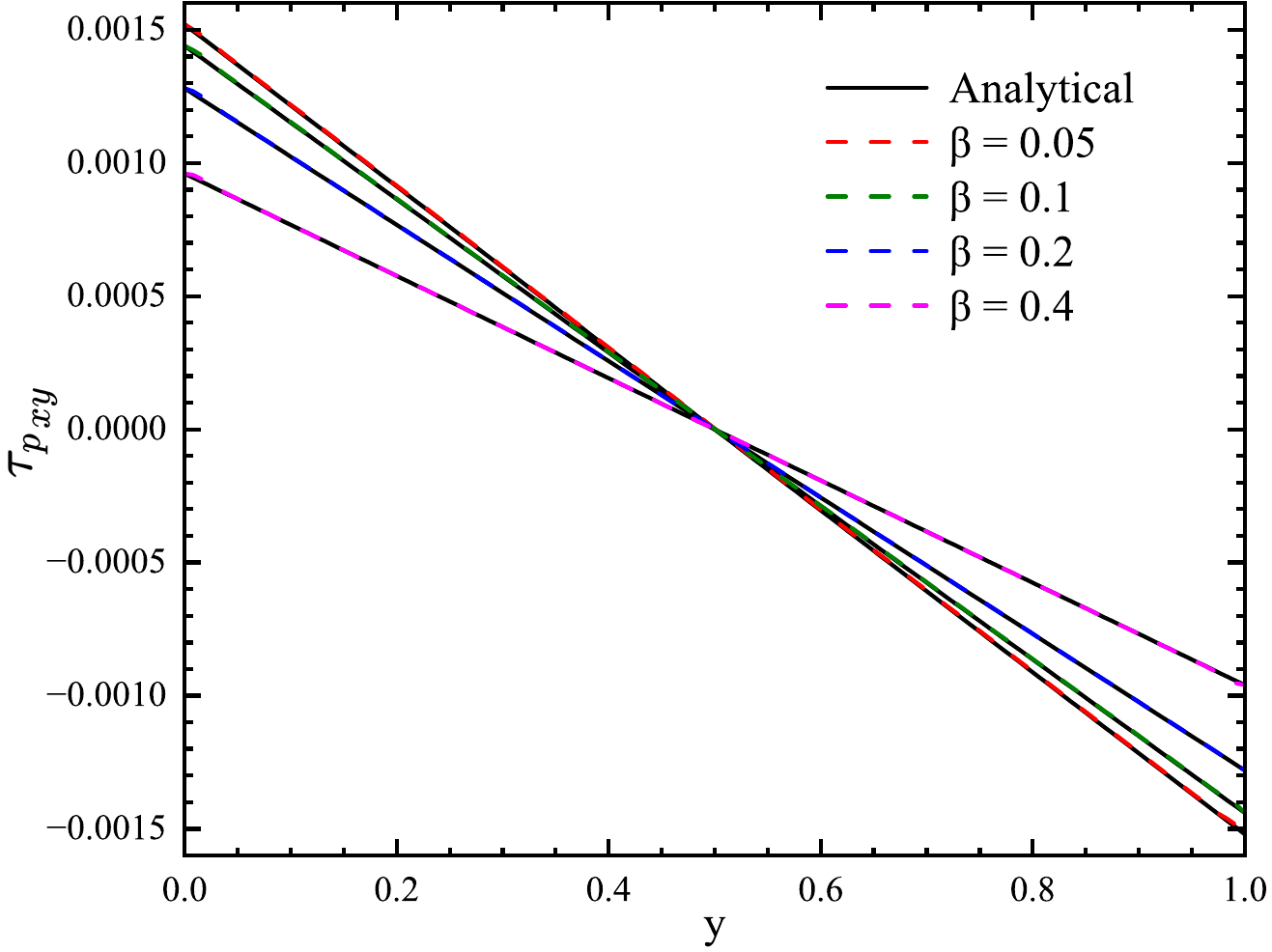} 
	\put(-230 ,168){(\textit{b})}
	\caption{The comparison of analytical and numerical results of (a) the normal stress component ${\tau_{{p_{xx}}}}$ and (b) tangential stress component ${\tau_{{p_{xy}}}}$.}
	\label{beta_tau_comparison}
\end{figure}

Fig. \ref{beta_Uc_transient} illustrates the time evolution of the transient centerline velocity $u_{c}^{*}$ obtained at $y = L/2$ for various kinematic viscosity ratios, and the numerical results align well with the analytical solution (see Eq. (\ref{poi_transient_solution})), verifying the validity of the current model in simulating transient viscoelastic flow. It is clear that as the kinematic viscosity ratio $\beta$ decreases, the intensity of centerline velocity fluctuations increases, and the time required for the system to reach a steady state also increases \cite{Zhang_CF2025}.
The process of centerline velocity evolution over time can be explained in simple terms. Initially, the fluid accelerates from a stationary state due to a constant body force, causing the viscoelastic stress to increase as the fluid is stretched. The viscous and viscoelastic effects compete against each other, and a decrease in centerline velocity occurs when the viscoelastic effect is dominant, resulting in fluctuations in velocity \cite{Zou_JNNFM2014}. After a long period of acceleration and deceleration, the flow eventually reaches a steady state due to a combination of viscous and viscoelastic effects.
\begin{figure}[H]
	\centering
	\includegraphics[scale=0.36]{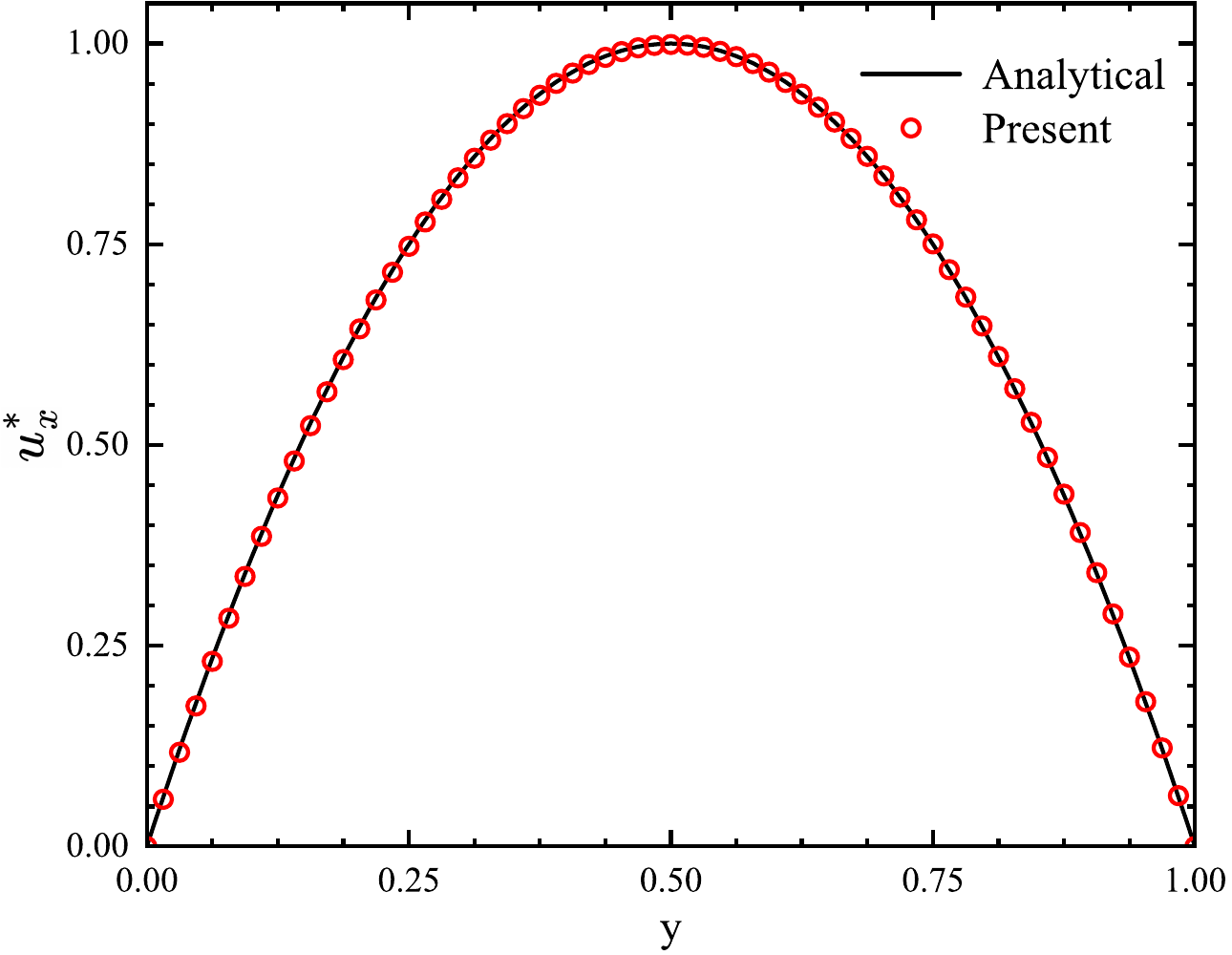} 
	\put(-230 ,170){(\textit{a})}
	\quad
	\includegraphics[scale=0.36]{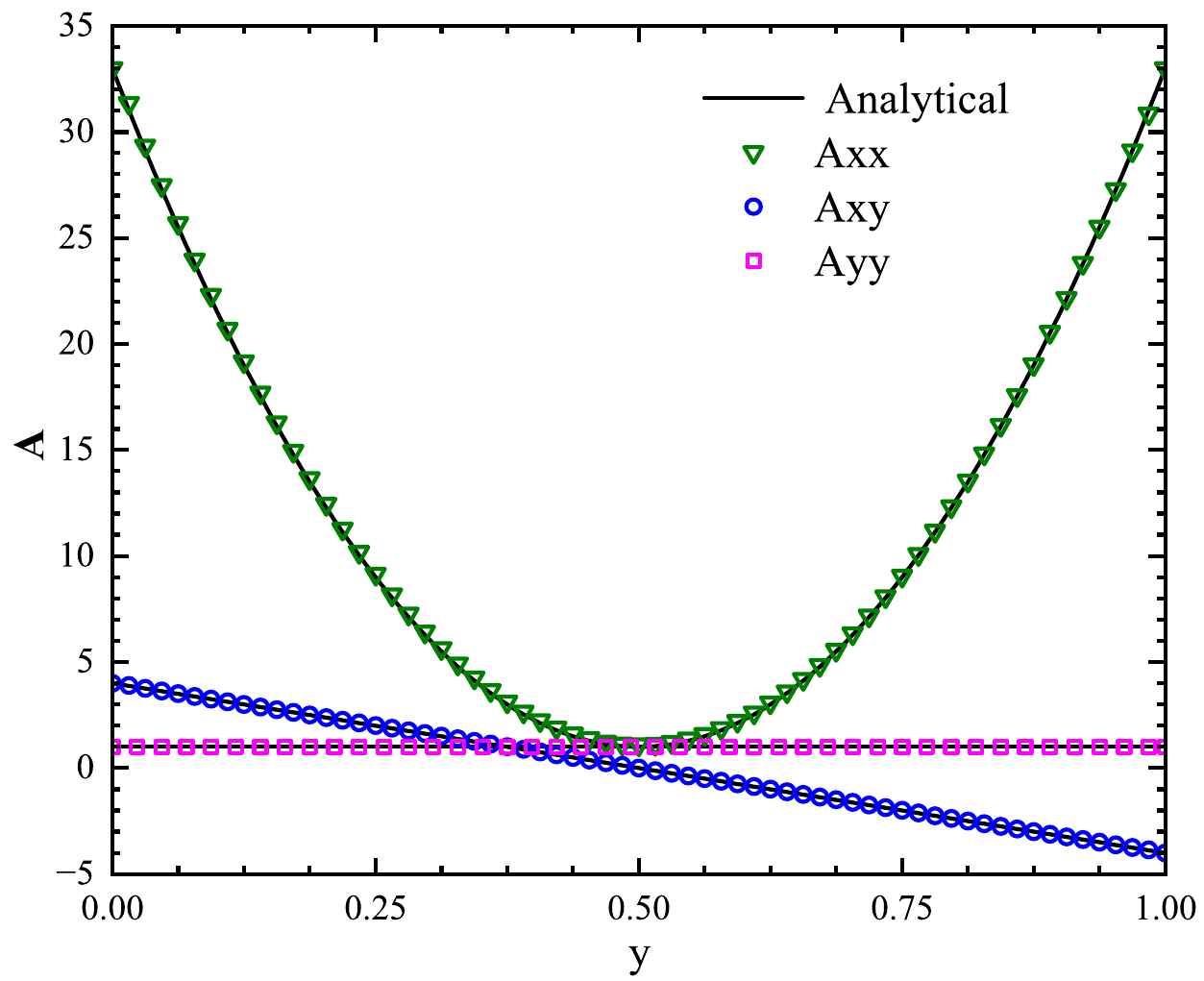} 
	\put(-225 ,170){(\textit{b})}
	\caption{The comparison of the analytical and numerical results of (a) the dimensionless velocity $u_{x}^{*}$ and (b) the components of the conformation tensor $\bf{A}$ at $x = L$ in steady state.}
	\label{Poi_steady_A_Uc_comparison}
\end{figure}
Fig. \ref{Poi_steady_A_Uc_comparison} illustrates the evolution of the dimensionless velocity $u_{x}^{*}$ and the components of the conformation tensor $\bf{A}$ with $y$ at $x = L$ when the flow reaches a steady state. We also present the numerical results of the stress tensor for various kinematic viscosity ratios and compare them to the stress tensor derived from the analytical conformation tensor solution, as shown in Fig. \ref{beta_tau_comparison}. The results indicate that the numerical solution closely aligns with the analytical solution, further validating that the current model can efficiently simulate the entire progression of viscoelastic flow.

Last, we also evaluate the performance of the present and original models with the lattice size of $256lu \times 128lu$, maintaining the iteration steps at 10000. The results in Table \ref{Poi_table} indicate that the present model not only decreases memory usage compared to the original one but also enhances computational efficiency, suggesting that it is more appropriate for large-scale simulations.
\begin{table}[htbp]
	\centering
	\caption{The comparison of the memory usage and computational time among the present and original schemes for 2D planar Poiseuille flow at the iteration step is 10000, in which ${\bf{A}}$ and ${\tau_p}$ represent the conformation tensor and polymer stress, respectively.}
	\begin{tabular}{cccc}  
		\toprule
		& Storage variables & Memory usage & Total time  \\
		\midrule
		Present & ${\bf{A}}[NX][NY][4]$, ${\tau_p}[NX][NY][4]$  & 94.58\% & 103.17$s$  \\
		Original & ${\bf{A}}[NX][NY][4]$, ${\tau_p}[NX][NY][4]$,  & 100\% & 115.48$s$  \\
		& $\frac{{\partial {\tau _p}}}{{\partial x}}[NX][NY][4]$, $\frac{{\partial {\tau _p}}}{{\partial y}}[NX][NY][4]$, & & \\
		\bottomrule
	\end{tabular}\label{Poi_table}
\end{table}

\subsection{Unsteady Womersley flow}

\begin{figure}[H]
	\centering
	\includegraphics[width=0.3\textwidth]{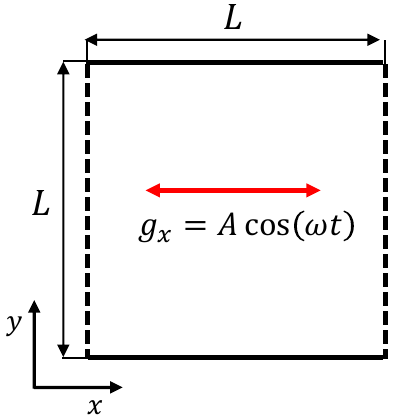}
	\caption{Schematic of Womersley flow.}
	\label{Womersley_flow}
\end{figure}

In this section, a series of Unsteady Womersley flow simulations are conducted using the present model. This classical problem is widely used to assess the numerical stability of both Newtonian fluids and viscoelastic fluid. The computational domain consists of a channel where a varying pressure gradient ${{\partial P} \mathord{\left/{\vphantom {{\partial P} {\partial y}}} \right.\kern-\nulldelimiterspace} {\partial x}} =  A\cos \left( {\omega t} \right)$ in the x-direction drives the fluid flow. Due to this pressure, the fluid in the channel displays a periodic oscillatory flow. In our simulations, the computational domain has a size of $200lu \times 200lu$. The periodic boundary condition is imposed on the inlet and outlet of the channel, while the no-slip condition is applied at the top and bottom walls. To simplify the implementation of the non-uniform pressure gradient, as shown in Fig. \ref{Womersley_flow}, an equivalent body force is defined as \cite{Zhang_CF2025}
\begin{equation}
	{g_x} =  A\cos \left( {\omega t} \right) = \frac{8}{{{L^2}}}\left( {{\nu _s} + {\nu _p}} \right){U}\cos \left( {\omega t} \right),
\end{equation}
in which $A$ and $\omega$ denote the amplitude and oscillation frequency, respectively. By incorporating the specified boundary conditions and pressure gradient into the macroscopic governing equations, Cosgrove et al. \cite{Cosgrove_JPA2003} derived an analytical solution for the velocity profile in a two-dimensional infinitely long channel. The analytical expression for the velocity is given by \cite{Zhang_CF2025}
\begin{equation}
	{u_x} = {\mathop{\rm Re}\nolimits} \left\{ {\frac{A}{{i\omega \rho }}\left[ {1 - \frac{{\cosh \left[ {\frac{1}{{\sqrt 2 }}\left( {1 + i} \right)\gamma \frac{{2y}}{L}} \right]}}{{\cosh \left[ {\frac{1}{{\sqrt 2 }}\left( {1 + i} \right)\gamma } \right]}}} \right]{e^{i\omega t}}} \right\},
\end{equation}
where $Re$ is the real part of the solution, $i$ denotes the imaginary unit. $\gamma  = {{L\sqrt {{\omega  \mathord{\left/{\vphantom {\omega  {{\nu _s}}}} \right.\kern-\nulldelimiterspace} {{\nu _s}}}} } \mathord{\left/{\vphantom {{L\sqrt {{\omega  \mathord{\left/{\vphantom {\omega  {{\nu _s}}}} \right.\kern-\nulldelimiterspace} {{\nu _s}}}} } 2}} \right.\kern-\nulldelimiterspace} 2}$ represents the Womersley parameter, which is a crucial quantity characterizing the Womersley flow.

\begin{figure}[H]
	\centering
	\includegraphics[scale=0.58]{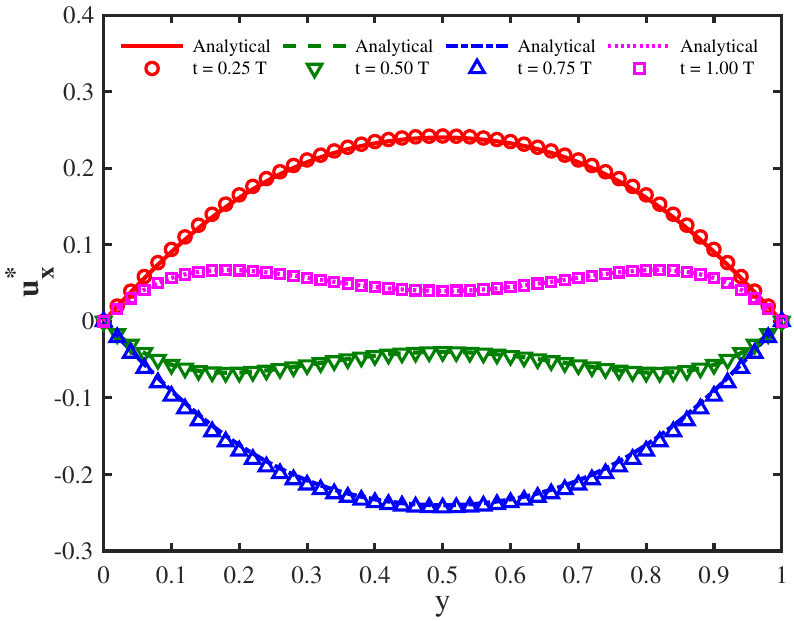} 
	\put(-230 ,168){(\textit{a})}
	\quad
	\includegraphics[scale=0.58]{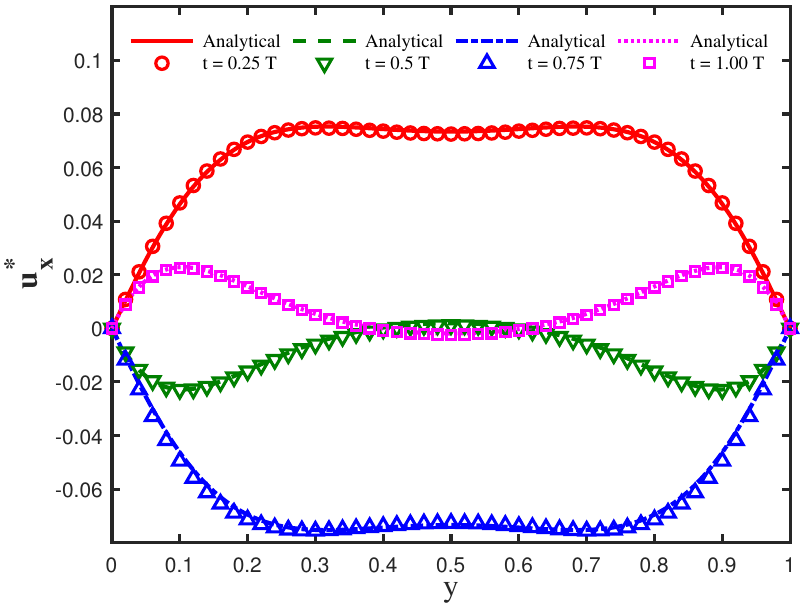} 
	\put(-230 ,168){(\textit{b})}
	\caption{The comparison of analytical and numerical results of the nondimensional velocity for (a) $Re=30$ and (b) $Re=90$.}
	\label{Womersley_Comparison}
\end{figure}

To validate the reliability of the present model, we first conducted a series of simulations for Newtonian fluid flow. The initial velocity field is set to zero, and the fluid is subjected to periodic oscillatory motion driven by a time-dependent external force. As shown in Fig. \ref{Womersley_Comparison}, after a sufficient time, we plot the dimensionless velocity $u_{x}^{*}$ as a function of $y$ at $x=L/2$ over one oscillation cycle for various Reynolds numbers and Womersley parameters. In Fig. \ref{Womersley_Comparison}(a), the parameters are set to $Re=30$ and $\gamma = 3.070$; in Fig. \ref{Womersley_Comparison}(b), $Re=90$ and $\gamma = 5.317$. The current results align well with the analytical solution.

\begin{figure}[H]
	\centering
	\includegraphics[scale=0.36]{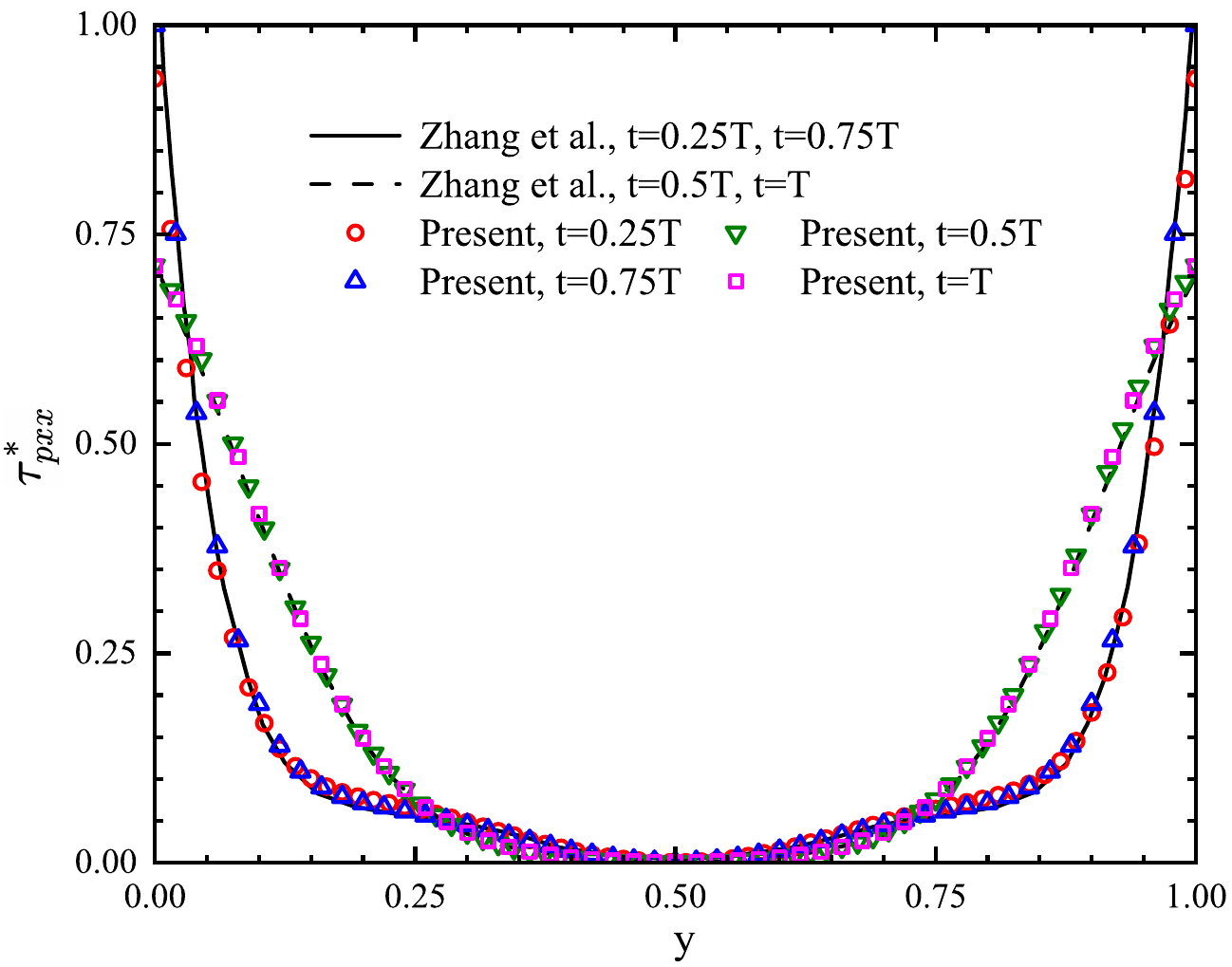} 
	\put(-230 ,168){(\textit{a})}
	\quad
	\includegraphics[scale=0.36]{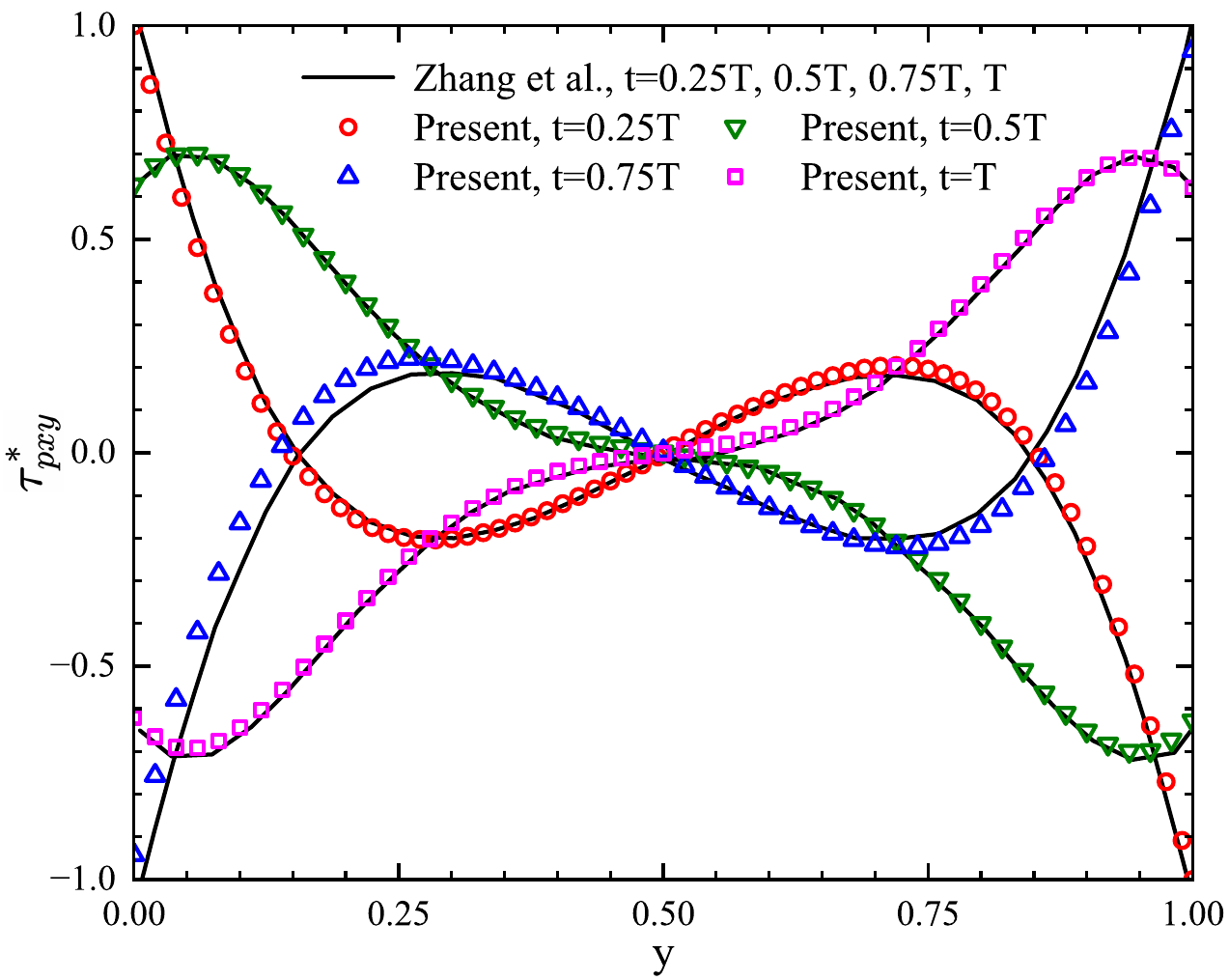} 
	\put(-230 ,168){(\textit{b})}
	\caption{The comparison of the dimensionless normal stress component ${\tau_{{p_{xx}}}}$ (a) and tangential stress component ${\tau_{{p_{xy}}}}$ (b) for $Re=30$, $Wi=5$, and $\beta=0.5$.}
	\label{Womersley_tau_p_Comparison}
\end{figure}

Additionally, we utilized the Oldroyd-B model to simulate a range of viscoelastic flows under the previously mentioned periodic external force conditions. To verify the accuracy of the current model in simulating viscoelastic flow, we adopted the same parameters as those used by Zhang et al. \cite{Zhang_CF2025}, with $Re=30$, $Wi=5$, and $\beta = 0.5$. Fig. \ref{Womersley_tau_p_Comparison}(a) and \ref{Womersley_tau_p_Comparison}(b) show the variation of the dimensionless normal and tangential stress components along the y-direction at $x=L/2$ over one oscillation cycle. The results demonstrate excellent agreement with the reference data, confirming the accuracy of the present model.

\subsection{3D Taylor–Green vortex}

\begin{figure}[H]
	\centering
	\includegraphics[scale=0.36]{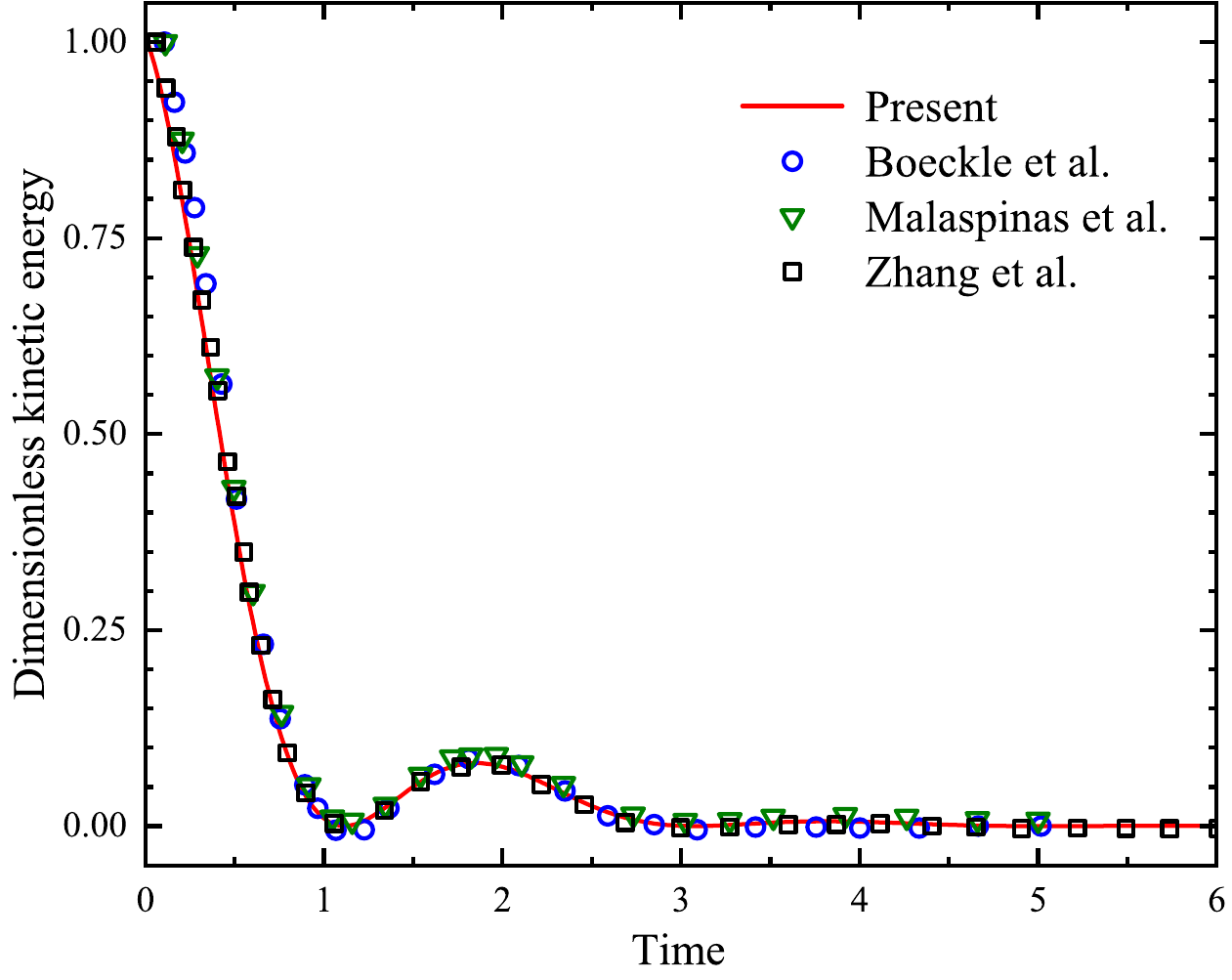} 
	\put(-220 ,168){(\textit{a})}
	\quad \quad
	\includegraphics[scale=0.36]{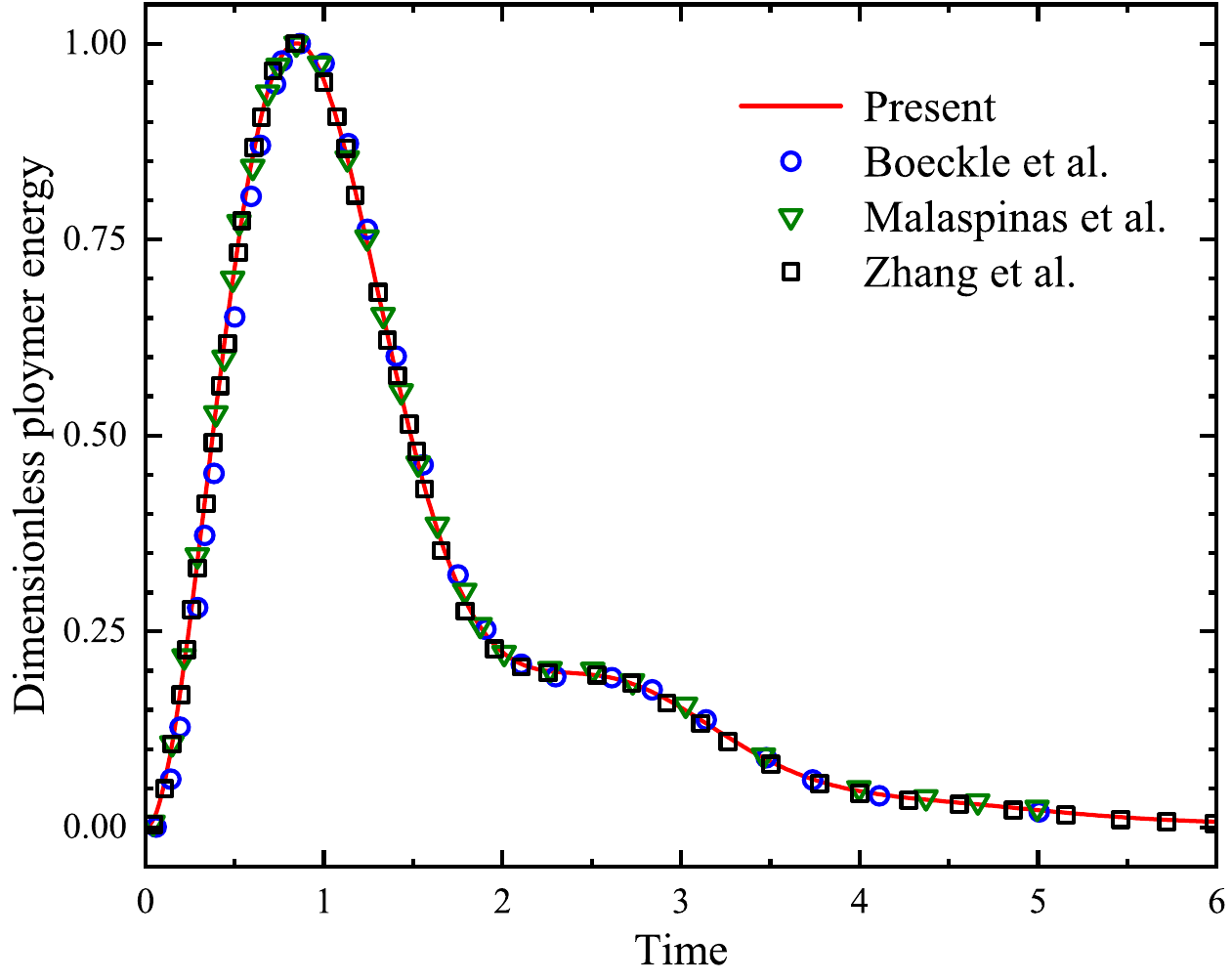} 
	\put(-220 ,168){(\textit{b})}
	\caption{Time evolution of the dimensionless kinetic energy (a) and dimensionless polymer energy (b) for $Wi=1.0$.}
	\label{Wi_1_Energy}
\end{figure}

To further validate that the proposed model can effectively simulate more complex viscoelastic flows, we conducted several simulations of 3D Taylor-Green vortex (TGV) using the Oldroyd-B model. The physical domain has a size of $\left[ {0,2\pi } \right] \times \left[ {0,2\pi } \right] \times \left[ {0,2\pi } \right]$, and the initial velocity field is described by \cite{Brachet_JFM1983}
\begin{equation}
	\begin{aligned}
		& u_x(\mathbf{x})=\frac{2 U_0}{\sqrt{3}} \sin \left(\theta+\frac{2 \pi}{3}\right) \sin (x) \cos (y) \cos (z), \\
		& u_y(\mathbf{x})=\frac{2 U_0}{\sqrt{3}} \sin \left(\theta-\frac{2 \pi}{3}\right) \cos (x) \sin (y) \cos (z) ,\\
		& u_z(\mathbf{x})=\frac{2 U_0}{\sqrt{3}} \sin (\theta) \cos (x) \cos (y) \sin (z),
	\end{aligned}
\end{equation}
in which $U_0$ represents the characteristic velocity in simulations, and $\theta=\pi/2$ denotes a parameter related to the initial vortex shape \cite{Zhang_CF2025}. The time-dependent average kinetic energy $KE\left( t \right)$ and polymer elastic energy $PE\left( t \right)$ are calculated by 
\begin{equation}
	KE\left( t \right) = \frac{1}{M}\sum\limits_i {\frac{{{\bf{u}}{{\left( {{x_i},t} \right)}^2}}}{2}} ,\quad{\rm{ }}PE\left( t \right) = \frac{1}{M}\sum\limits_i {\frac{{\rm{tr}\left( {{\tau _p}\left( {{x_i},t} \right)} \right)}}{2}} ,
\end{equation}
in which $M$ is the total number of the grid point. Another point to note is that the initialization of the pressure field significantly affects the results. In our simulations, the initialized pressure field is given by \cite{Zhang_CF2025}
\begin{equation}
		\begin{aligned}
			p(\mathbf{x})= & p_0+ U_0^2\left(\frac{1}{12}[\cos (2 x) \cos (2 y)+2 \cos (2 z)]\right. \\
			& \left.+\frac{1}{48}[\cos (2 x) \cos (2 z)+2 \cos (2 y)]+\frac{1}{48}[\cos (2 y) \cos (2 z)+2 \cos (2 x)]\right),
		\end{aligned}
\end{equation}
in which $p_0$ is the reference pressure of this case, set to $p_0=0.0$. The numerical results from the current model are compared with those obtained by Boeckle et al. utilizing a highly accurate Fourier pseudospectral method \cite{Boeckle_2009}, Malaspinas et al. employing a LB method \cite{Malaspinas_JNNFM2010}, and Zhang et al. using a viscoelastic lattice Boltzmann flux solver \cite{Zhang_CF2025}. In our simulations, the whole computational domain has a size of $100lu \times 100lu \times 100lu$, and the periodic boundary conditions are applied to all boundaries. The initial viscoelastic stress tensor is taken equal to zero, and the dimensionless parameters are set to $Re=1.0$, $Wi=1.0, 10.0$, $\beta=0.1$. The characteristic velocity and length are specified as $U_0=0.02$ and $L=1.0$, respectively. Fig. \ref{Wi_1_Energy} illustrates the evolution of the dimensionless kinetic and polymer energies for $Wi=1.0$ over time. The dimensionless kinetic energy is calculated by dividing the kinetic energy by its initial value, while the dimensionless polymer energy is derived by dividing the polymer energy by its maximum value. The whole energy evolution can be explained as follows: due to the given initial conditions, the fluid has an initial kinetic energy, and the polymer elastic potential energy is zero. Under the influence of the initial kinetic energy, the polymer molecules start to stretch during the deformation of the fluid, and a part of the kinetic energy is converted into polymer elastic potential energy ($0<t<1$). Subsequently, when the velocity gradient is insufficient to continue stretching the polymer molecules, the elastic potential energy stored in the polymer is partially converted back into kinetic energy ($t=1$). This results in a local peak in the kinetic energy ($t=2$). This process repeats until the total energy of the system disappears through viscous dissipation. 
\begin{figure}[H]
	\centering
	\includegraphics[scale=0.33]{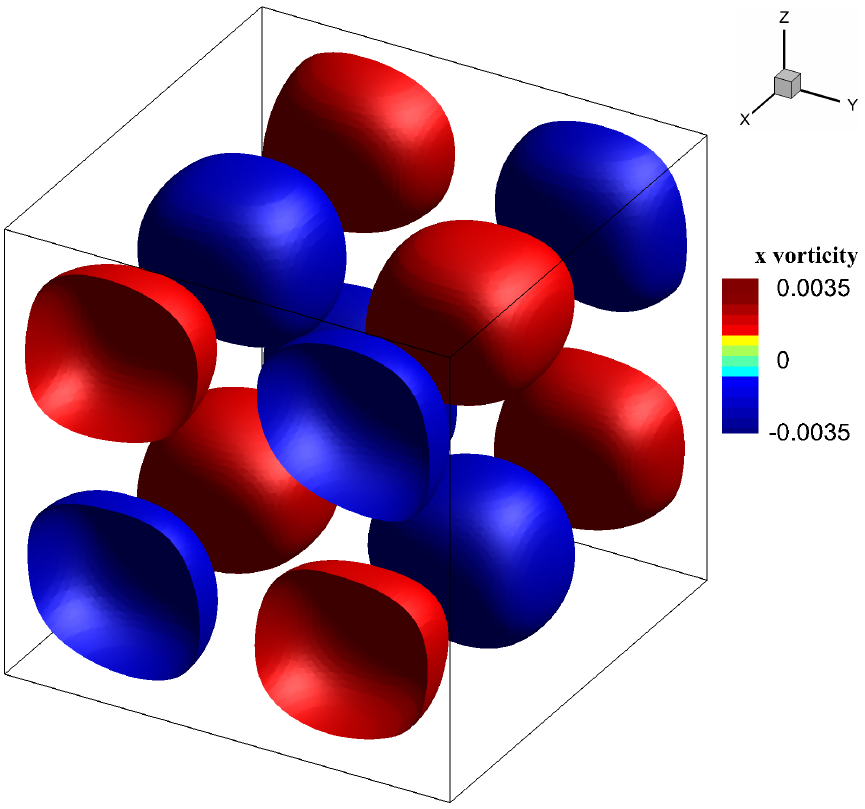} 
	\put(-120 ,-10){(\textit{a}) $Wi=1.0$, $t=0.5$}
	\quad 
	\includegraphics[scale=0.33]{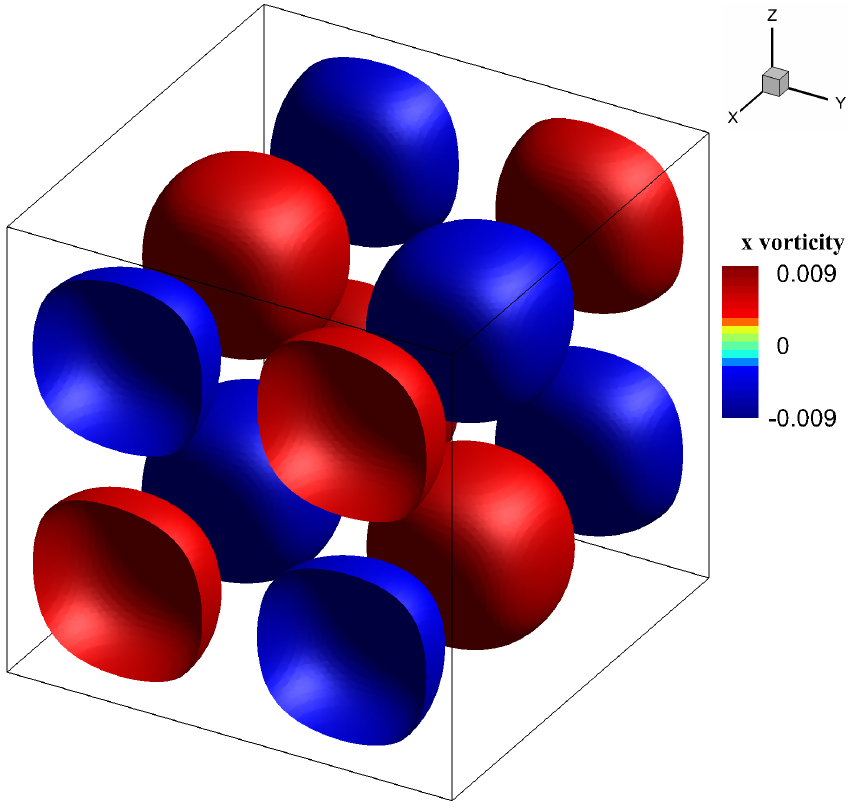} 
	\put(-120 ,-10){(\textit{b}) $Wi=1.0$, $t=1.0$}
	\quad 
	\includegraphics[scale=0.33]{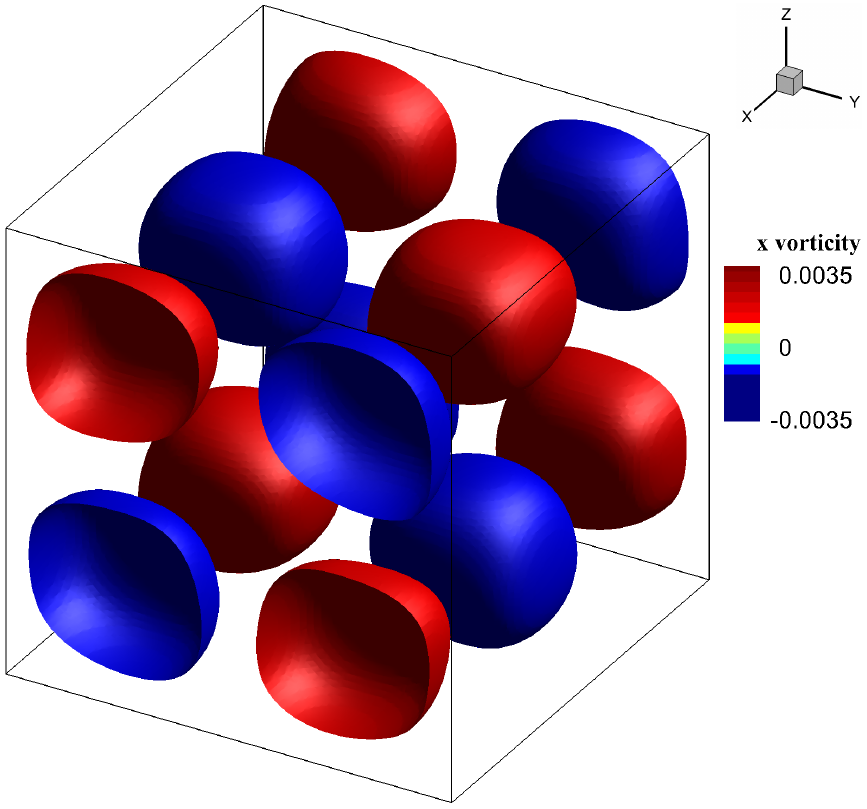} 
	\put(-120 ,-10){(\textit{c}) $Wi=1.0$, $t=2.0$}
	\caption{Iso-surfaces of the x-component of vorticity for Wi = 1.0 at (a) $t=0.5$, (b) $t=1.0$, and (c) $t=2.0$.}
	\label{Wi_1_evolution}
\end{figure}

\begin{figure}[H]
	\centering
	\includegraphics[scale=0.36]{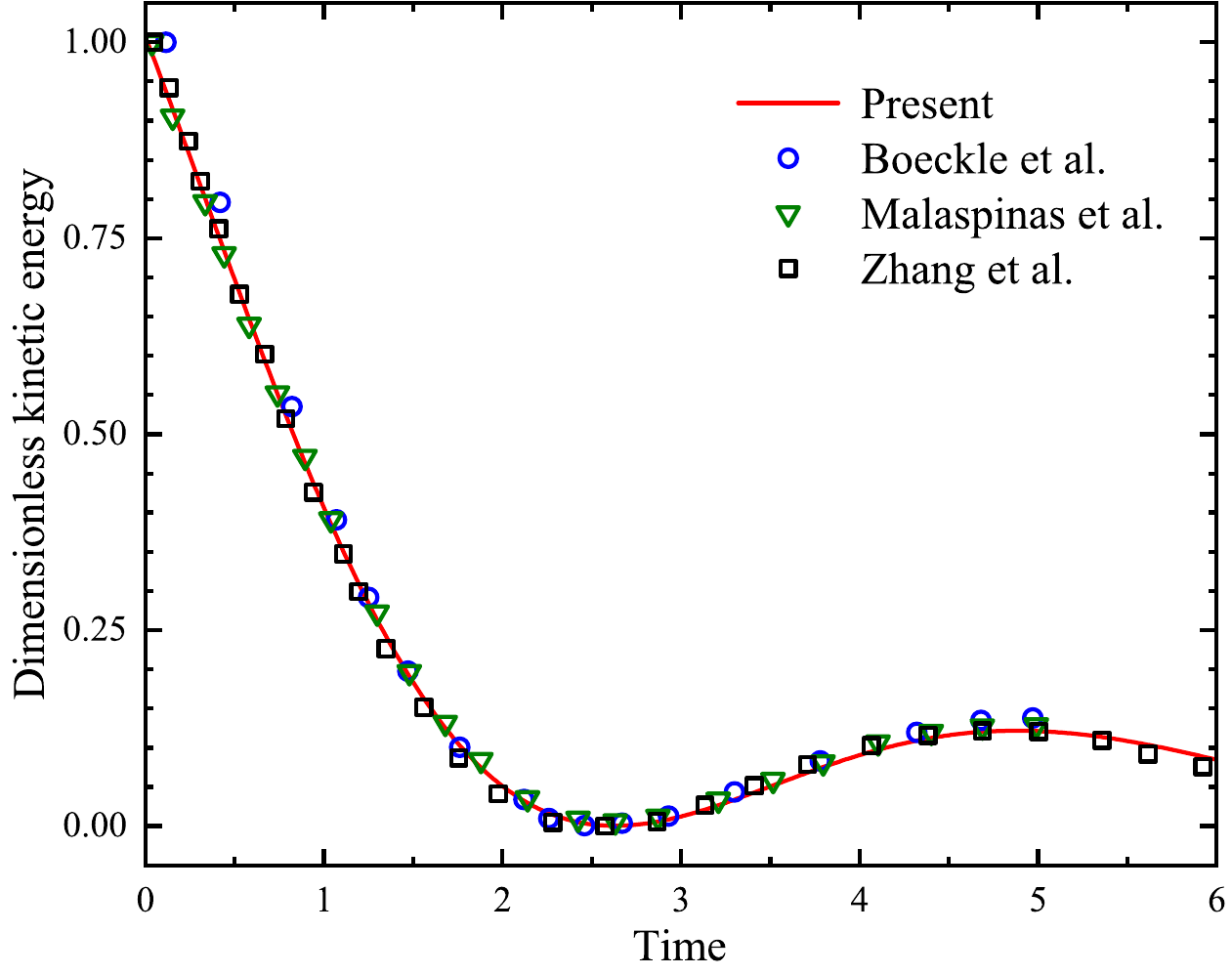} 
	\put(-220 ,168){(\textit{a})}
	\quad \quad
	\includegraphics[scale=0.36]{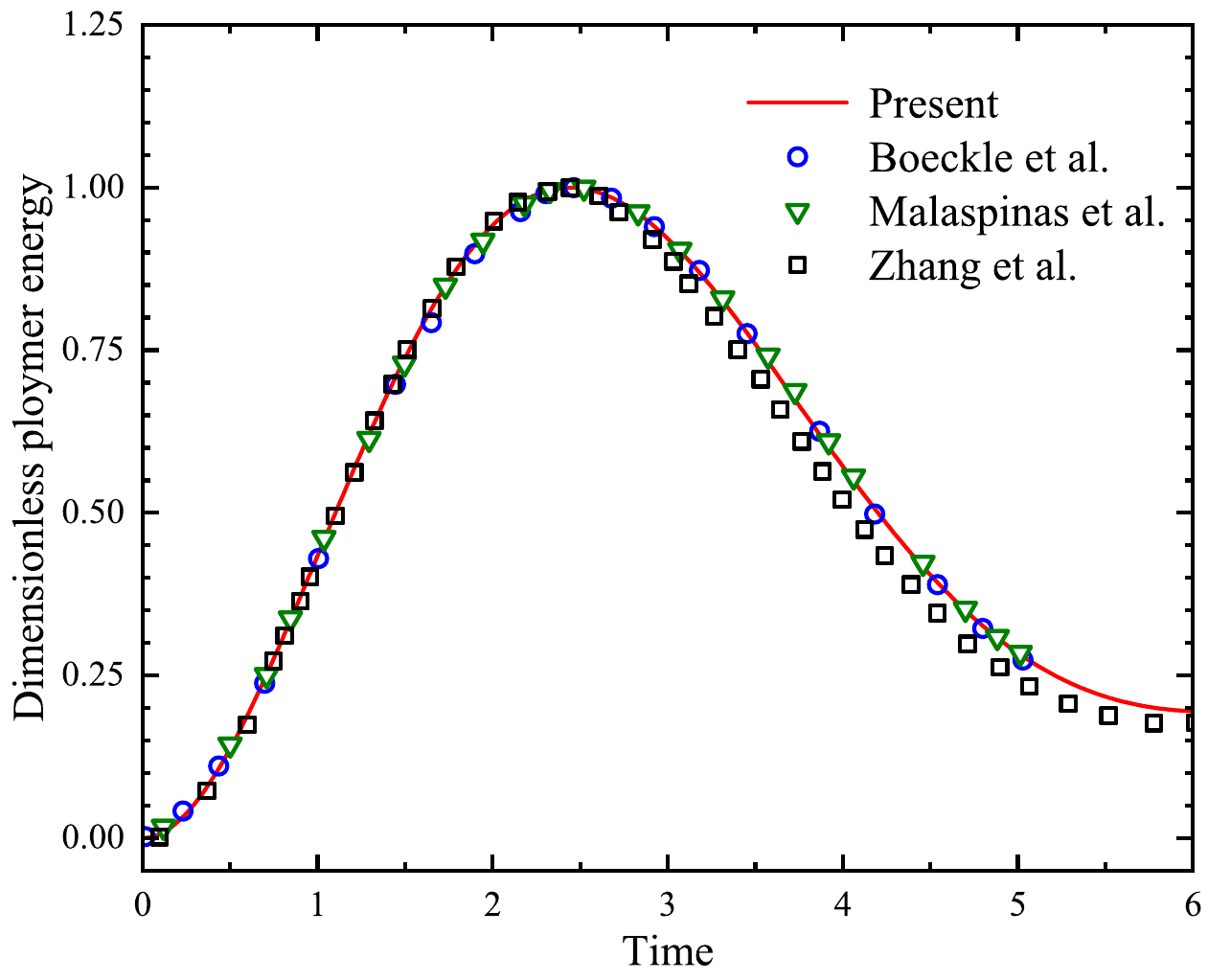} 
	\put(-220 ,168){(\textit{b})}
	\caption{Time evolution of the dimensionless kinetic energy (a) and dimensionless polymer energy (b) for $Wi=10.0$.}
	\label{Wi_10_Energy}
\end{figure}

\begin{figure}[H]
	\centering
	\includegraphics[scale=0.33]{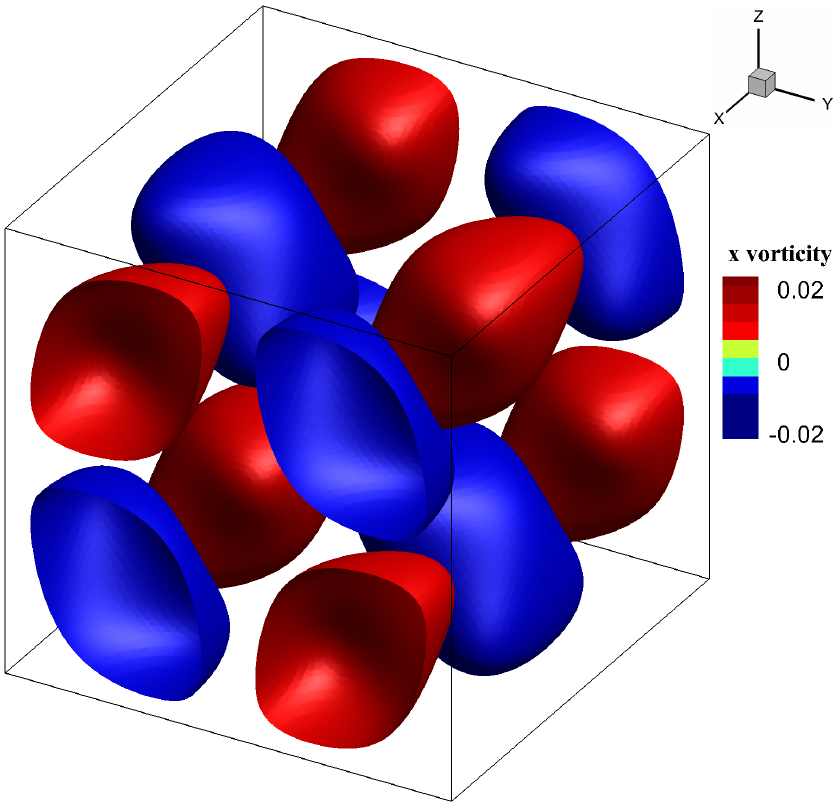} 
	\put(-120 ,-10){(\textit{a}) $Wi=10.0$, $t=1$}
	\quad
	\includegraphics[scale=0.33]{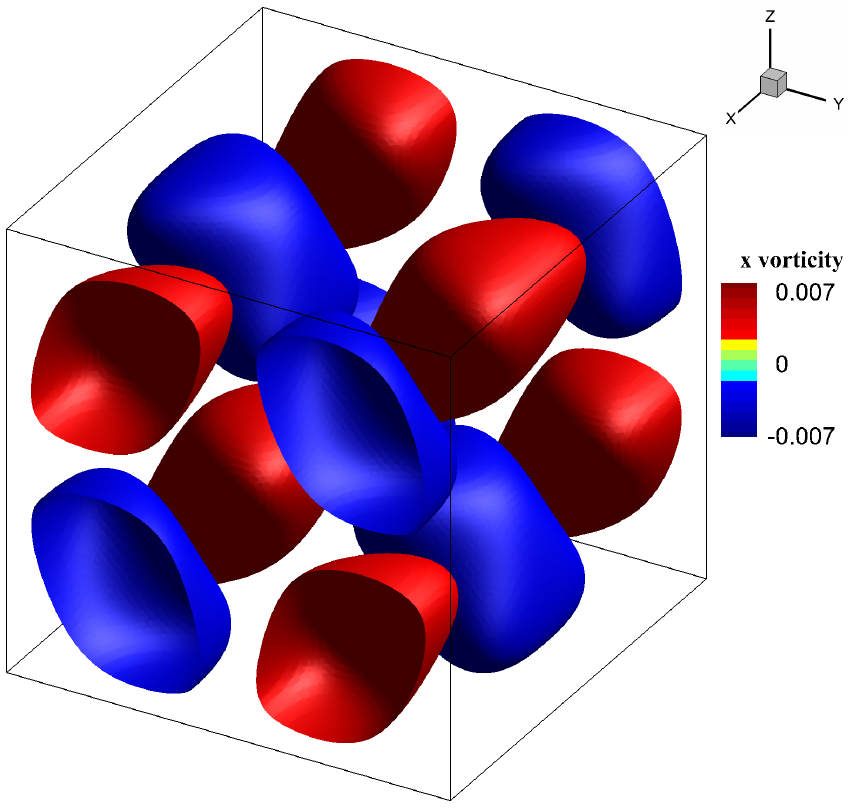} 
	\put(-120 ,-10){(\textit{b}) $Wi=10.0$, $t=2$}
	\quad
	\includegraphics[scale=0.33]{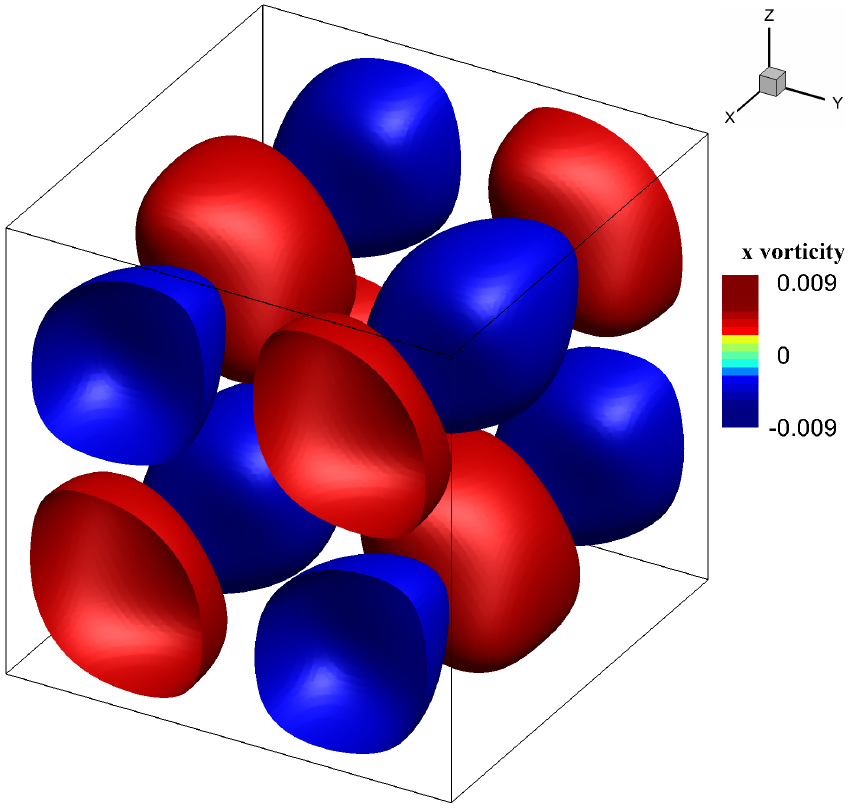} 
	\put(-120 ,-10){(\textit{c}) $Wi=10.0$, $t=4$}
	\caption{Iso-surfaces of the x-component of vorticity for Wi = 10.0 at (a) $t=1.0$, (b) $t=2.0$, and (c) $t=4.0$.}
	\label{Wi_10_evolution}
\end{figure}

Fig. \ref{Wi_1_evolution} illustrates the iso-surfaces of the vorticity x-component at various dimensionless times, and Figs. \ref{Wi_1_evolution} (b) and (c) show the reversal of vortex rotation in the TGV due to viscoelastic effects \cite{Zhang_CF2025}. Additionally, we also plot the time evolution of the dimensionless kinetic and elastic energies for $Wi = 10.0$, as illustrated in Fig. \ref{Wi_10_Energy}. It is clear that the current results strongly align with those obtained using a high-accuracy spectral method \cite{Boeckle_2009}. As the Weissenberg number increases, the time interval between two peaks of kinetic energy lengthens because a high Weissenberg number indicates a longer relaxation time for the polymer \cite{Malaspinas_JNNFM2010}. Moreover, Fig. \ref{Wi_10_evolution} illustrates the iso-surface of the x-component of vorticity for different dimensionless times, and similar to the case of $Wi=1.0$, a reversal of the vortex rotation also occurs. 

Furthermore, we test the performance of both the present and original models in the three-dimensional TGV problem, using fixed iteration steps of 10000 with the lattice size of $100lu \times 100lu \times 100lu$. The results presented in Table \ref{3D_Taylor_green_table} demonstrate that the present model offers significant advantages under three-dimensional condition, not only by reducing memory occupancy but also by enhancing computational efficiency. These findings indicate its strong potential for application in large-scale simulations under three-dimensional condition.

\begin{table}[htbp]
	\centering
	\caption{The comparison of the memory usage and computational time among the present and original schemes for 3D Taylor-Green vortex at the iteration step is 10000, in which ${\bf{A}}$ and ${\tau_p}$ represent the conformation tensor and polymer stress, respectively.}
	\begin{tabular}{cccc}  
		\toprule
		& Storage variables & Memory usage & Total time  \\
		\midrule
		Present & ${\bf{A}}[NX][NY][NZ][9]$, ${\tau_p}[NX][NY][NZ][9]$  & 96.98\% & 1825.56$s$  \\
		Original & ${\bf{A}}[NX][NY][NZ][9]$, ${\tau_p}[NX][NY][NZ][9]$,  & 100\% & 2150.08$s$  \\
		& $\frac{{\partial {\tau _p}}}{{\partial x}}[NX][NY][NZ][9]$, $\frac{{\partial {\tau _p}}}{{\partial y}}[NX][NY][NZ][9]$, $\frac{{\partial {\tau _p}}}{{\partial z}}[NX][NY][NZ][9]$ & & \\
		\bottomrule
	\end{tabular}\label{3D_Taylor_green_table}
\end{table}

\section{Conclusion}\label{section_4}

In this work, we develop a fluctuating LB method for simulating viscoelastic fluid flows characterized by the Oldroyd-B model. By introducing a polymer stress fluctuation term into the velocity evolution equation, the proposed scheme successfully avoids the direct computation of the polymer stress tensor divergence.

The developed model is validated through a series of benchmark tests, including planar Poiseuille flow, simplified four-roll mill, unsteady Womersley flow, and the three-dimensional Taylor-Green vortex. The results demonstrate that the proposed model effectively captures key phenomena inherent to viscoelastic fluid flows. These phenomena include the competition between viscous and elastic forces in planar Poiseuille flow, the formation of local secondary flows in the four-roll mill, and the reversal of vortex rotation in the TGV. Moreover, the results of the present model are in excellent agreement with both analytical solutions and results from previous numerical studies, confirming its reliability in simulating viscoelastic fluid dynamics.

Furthermore, performance analysis also reveals that the present model reduces memory usage and enhances computational efficiency. These improvements demonstrate the current model's potential for large-scale simulations of complex viscoelastic systems, making it as a valuable tool for future research in the field. Future work will extend this framework to more general constitutive models and explore its application in multiphase and non-isothermal viscoelastic fluid flows.

\section*{Appendix A. The computation of ${f_i}^{B}$ using a pseudo-inverse method}

In this appendix, we give the calculation process for the polymer stress fluctuation term ${f_i}^B$. Using Eqs. (\ref{zero_order_fB}), (\ref{first_order_fB}), and (\ref{second_order_fB}), and with $c=1$, the corresponding system of linear equations can be written as ${\bf{C}}{{\bf{f}}^{\left( B \right)}} = {\bf{b}}$, in which
\begin{subequations}
	\begin{gather}
		{\bf{C}} = \left( {\begin{array}{*{20}{c}}
				0&0&0&0&0&1&{ - 1}&1&{ - 1}\\
				0&1&0&1&0&1&1&1&1\\
				0&0&0&0&0&1&{ - 1}&1&{ - 1}\\
				0&0&1&0&1&1&1&1&1\\
				1&1&1&1&1&1&1&1&1\\
				0&1&0&{ - 1}&0&1&{ - 1}&{ - 1}&1\\
				0&0&1&0&{ - 1}&1&1&{ - 1}&{ - 1}
		\end{array}} \right),\\
		{\bf{b}} = {\left( { - \frac{{{\tau _{{p_{yx}}}}}}{{{\tau _f}}}, - \frac{{{\tau _{{p_{xx}}}}}}{{{\tau _f}}}, - \frac{{{\tau _{{p_{xy}}}}}}{{{\tau _f}}}, - \frac{{{\tau _{{p_{yy}}}}}}{{{\tau _f}}},0,0,0} \right)^{\rm{T}}},\\
		{{\bf{f}}^{\left( B \right)}} = {\left( {f_0^B,f_1^B,f_2^B,f_3^B,f_4^B,f_5^B,f_6^B,f_7^B,f_8^B} \right)^{\rm{T}}}.
	\end{gather}
\end{subequations}

According to the formula for the calculation of the pseudo-inverse, the pseudo-inverse matrix of ${\bf{C}}$ can be expressed as
\begin{equation}
	{{\bf{C}}^{ - 1}} = {{\bf{C}}^{\rm{T}}}{\left( {{\bf{C}}{{\bf{C}}^{\rm{T}}}} \right)^{ - 1}} = \frac{1}{{72}}\left( {\begin{array}{*{20}{c}}
			0&{ - 24}&0&{ - 24}&{40}&0&0\\
			0&{12}&0&{ - 24}&{16}&{12}&0\\
			0&{ - 24}&0&{12}&{16}&0&{12}\\
			0&{12}&0&{ - 24}&{16}&{ - 12}&0\\
			0&{ - 24}&0&{12}&{16}&0&{ - 12}\\
			9&{12}&9&{12}&{ - 8}&{12}&{12}\\
			{ - 9}&{12}&{ - 9}&{12}&{ - 8}&{ - 12}&{12}\\
			9&{12}&9&{12}&{ - 8}&{ - 12}&{ - 12}\\
			{ - 9}&{12}&{ - 9}&{12}&{ - 8}&{12}&{ - 12}
	\end{array}} \right),
\end{equation}
in which ${{\bf{C}}^{ - 1}}$ and ${{\bf{C}}^{\rm{T}}}$ represent the inverse and transport of matrix ${\bf{C}}$, respectively. Then, ${{\bf{f}}^{\left( B \right)}}$ ($c=1$) can be calculated by
\begin{equation}
	{{\bf{f}}^{\left( B \right)}} = {{\bf{C}}^{ - 1}}{\bf{b}} = \frac{1}{{72}}\left( {\begin{array}{*{20}{c}}
			0&{ - 24}&0&{ - 24}&{40}&0&0\\
			0&{12}&0&{ - 24}&{16}&{12}&0\\
			0&{ - 24}&0&{12}&{16}&0&{12}\\
			0&{12}&0&{ - 24}&{16}&{ - 12}&0\\
			0&{ - 24}&0&{12}&{16}&0&{ - 12}\\
			9&{12}&9&{12}&{ - 8}&{12}&{12}\\
			{ - 9}&{12}&{ - 9}&{12}&{ - 8}&{ - 12}&{12}\\
			9&{12}&9&{12}&{ - 8}&{ - 12}&{ - 12}\\
			{ - 9}&{12}&{ - 9}&{12}&{ - 8}&{12}&{ - 12}
	\end{array}} \right)\left( {\begin{array}{*{20}{c}}
			{ - \frac{{{\tau _{{p_{yx}}}}}}{{{\tau _f}}}}\\
			{ - \frac{{{\tau _{{p_{xx}}}}}}{{{\tau _f}}}}\\
			{ - \frac{{{\tau _{{p_{xy}}}}}}{{{\tau _f}}}}\\
			{ - \frac{{{\tau _{{p_{yy}}}}}}{{{\tau _f}}}}\\
			0\\
			0\\
			0
	\end{array}} \right) =  \frac{-1}{{72 {\tau _f}}}\left( {\begin{array}{*{20}{c}}
			0&{ - 24}&0&{ - 24}\\
			0&{12}&0&{ - 24}\\
			0&{ - 24}&0&{12}\\
			0&{12}&0&{ - 24}\\
			0&{ - 24}&0&{12}\\
			9&{12}&9&{12}\\
			{ - 9}&{12}&{ - 9}&{12}\\
			9&{12}&9&{12}\\
			{ - 9}&{12}&{ - 9}&{12}
	\end{array}} \right)\left( {\begin{array}{*{20}{c}}
			{{\tau _{{p_{yx}}}}}\\
			{{\tau _{{p_{xx}}}}}\\
			{{\tau _{{p_{xy}}}}}\\
			{{\tau _{{p_{yy}}}}}
	\end{array}} \right).
\end{equation}

For the three-dimensional case, we utilize the $\rm{D3Q19}$ model, and its corresponding discrete velocities $\mathbf{c}_i$ can be expressed as
\begin{equation}
	\mathbf{c}_i = c \left[\begin{array}{lllllllllllllllllll}
		0 & 1 & -1 & 0 & 0 & 0 & 0 & 1 & -1 & 1 & -1 & 1 & -1 & 1 & -1 & 0 & 0 & 0 & 0 \\
		0 & 0 & 0 & 1 & -1 & 0 & 0 & 1 & -1 & -1 & 1 & 0 & 0 & 0 & 0 & 1 & -1 & 1 & -1 \\
		0 & 0 & 0 & 0 & 0 & 1 & -1 & 0 & 0 & 0 & 0 & 1 & -1 & -1 & 1 & 1 & -1 & -1 & 1
	\end{array}\right],
\end{equation}
the polymer stress fluctuation term ${{\bf{f}}^{\left( B \right)}}$ (in the case of $c=1$) is calculated by
\begin{equation}
	\left( \begin{array}{l}
		f_0^B\\
		f_1^B\\
		f_2^B\\
		f_3^B\\
		f_4^B\\
		f_5^B\\
		f_6^B\\
		f_7^B\\
		f_8^B\\
		f_9^B\\
		f_{10}^B\\
		f_{11}^B\\
		f_{12}^B\\
		f_{13}^B\\
		f_{14}^B\\
		f_{15}^B\\
		f_{16}^B\\
		f_{17}^B\\
		f_{18}^B
	\end{array} \right) = \left( {\begin{array}{*{20}{c}}
			{ - \frac{5}{{21}}}&{ - \frac{5}{{21}}}&{ - \frac{5}{{21}}}&0&0&0&0&0&0&0&0&0&{\frac{3}{7}}\\
			{\frac{1}{{42}}}&{ - \frac{1}{7}}&{ - \frac{1}{7}}&0&0&0&0&0&0&{\frac{1}{{10}}}&0&0&{\frac{4}{{21}}}\\
			{\frac{1}{{42}}}&{ - \frac{1}{7}}&{ - \frac{1}{7}}&0&0&0&0&0&0&{ - \frac{1}{{10}}}&0&0&{\frac{4}{{21}}}\\
			{ - \frac{1}{7}}&{\frac{1}{{42}}}&{ - \frac{1}{7}}&0&0&0&0&0&0&0&{\frac{1}{{10}}}&0&{\frac{4}{{21}}}\\
			{ - \frac{1}{7}}&{\frac{1}{{42}}}&{ - \frac{1}{7}}&0&0&0&0&0&0&0&{ - \frac{1}{{10}}}&0&{\frac{4}{{21}}}\\
			{ - \frac{1}{7}}&{ - \frac{1}{7}}&{\frac{1}{{42}}}&0&0&0&0&0&0&0&0&{\frac{1}{{10}}}&{\frac{4}{{21}}}\\
			{ - \frac{1}{7}}&{ - \frac{1}{7}}&{\frac{1}{{42}}}&0&0&0&0&0&0&0&0&{ - \frac{1}{{10}}}&{\frac{4}{{21}}}\\
			{\frac{5}{{42}}}&{\frac{5}{{42}}}&{ - \frac{1}{{21}}}&{\frac{1}{8}}&0&{\frac{1}{8}}&0&0&0&{\frac{1}{{10}}}&{\frac{1}{{10}}}&0&{ - \frac{1}{{21}}}\\
			{\frac{5}{{42}}}&{\frac{5}{{42}}}&{ - \frac{1}{{21}}}&{\frac{1}{8}}&0&{\frac{1}{8}}&0&0&0&{ - \frac{1}{{10}}}&{ - \frac{1}{{10}}}&0&{ - \frac{1}{{21}}}\\
			{\frac{5}{{42}}}&{\frac{5}{{42}}}&{ - \frac{1}{{21}}}&{ - \frac{1}{8}}&0&{ - \frac{1}{8}}&0&0&0&{\frac{1}{{10}}}&{ - \frac{1}{{10}}}&0&{ - \frac{1}{{21}}}\\
			{\frac{5}{{42}}}&{\frac{5}{{42}}}&{ - \frac{1}{{21}}}&{ - \frac{1}{8}}&0&{ - \frac{1}{8}}&0&0&0&{ - \frac{1}{{10}}}&{\frac{1}{{10}}}&0&{ - \frac{1}{{21}}}\\
			{\frac{5}{{42}}}&{ - \frac{1}{{21}}}&{\frac{5}{{42}}}&0&{\frac{1}{8}}&0&0&{\frac{1}{8}}&0&{\frac{1}{{10}}}&0&{\frac{1}{{10}}}&{ - \frac{1}{{21}}}\\
			{\frac{5}{{42}}}&{ - \frac{1}{{21}}}&{\frac{5}{{42}}}&0&{\frac{1}{8}}&0&0&{\frac{1}{8}}&0&{ - \frac{1}{{10}}}&0&{ - \frac{1}{{10}}}&{ - \frac{1}{{21}}}\\
			{\frac{5}{{42}}}&{ - \frac{1}{{21}}}&{\frac{5}{{42}}}&0&{ - \frac{1}{8}}&0&0&{ - \frac{1}{8}}&0&{\frac{1}{{10}}}&0&{ - \frac{1}{{10}}}&{ - \frac{1}{{21}}}\\
			{\frac{5}{{42}}}&{ - \frac{1}{{21}}}&{\frac{5}{{42}}}&0&{ - \frac{1}{8}}&0&0&{ - \frac{1}{8}}&0&{ - \frac{1}{{10}}}&0&{\frac{1}{{10}}}&{ - \frac{1}{{21}}}\\
			{ - \frac{1}{{21}}}&{\frac{5}{{42}}}&{\frac{5}{{42}}}&0&0&0&{\frac{1}{8}}&0&{\frac{1}{8}}&0&{\frac{1}{{10}}}&{\frac{1}{{10}}}&{ - \frac{1}{{21}}}\\
			{ - \frac{1}{{21}}}&{\frac{5}{{42}}}&{\frac{5}{{42}}}&0&0&0&{\frac{1}{8}}&0&{\frac{1}{8}}&0&{ - \frac{1}{{10}}}&{ - \frac{1}{{10}}}&{ - \frac{1}{{21}}}\\
			{ - \frac{1}{{21}}}&{\frac{5}{{42}}}&{\frac{5}{{42}}}&0&0&0&{ - \frac{1}{8}}&0&{ - \frac{1}{8}}&0&{\frac{1}{{10}}}&{ - \frac{1}{{10}}}&{ - \frac{1}{{21}}}\\
			{ - \frac{1}{{21}}}&{\frac{5}{{42}}}&{\frac{5}{{42}}}&0&0&0&{ - \frac{1}{8}}&0&{ - \frac{1}{8}}&0&{ - \frac{1}{{10}}}&{\frac{1}{{10}}}&{ - \frac{1}{{21}}}
	\end{array}} \right)\left( {\begin{array}{*{20}{c}}
			{ - \frac{{{\tau _{{p_{xx}}}}}}{{{\tau _f}}}}\\
			{ - \frac{{{\tau _{{p_{yy}}}}}}{{{\tau _f}}}}\\
			{ - \frac{{{\tau _{{p_{zz}}}}}}{{{\tau _f}}}}\\
			{ - \frac{{{\tau _{{p_{xy}}}}}}{{{\tau _f}}}}\\
			{ - \frac{{{\tau _{{p_{xz}}}}}}{{{\tau _f}}}}\\
			{ - \frac{{{\tau _{{p_{yx}}}}}}{{{\tau _f}}}}\\
			{ - \frac{{{\tau _{{p_{yz}}}}}}{{{\tau _f}}}}\\
			{ - \frac{{{\tau _{{p_{zx}}}}}}{{{\tau _f}}}}\\
			{ - \frac{{{\tau _{{p_{zy}}}}}}{{{\tau _f}}}}\\
			0\\
			0\\
			0\\
			0
	\end{array}} \right).
\end{equation}

\section*{Appendix B. The calculation of the pressure}\label{calculation_pressure}

In this appendix, the calculation formula for the pressure is given. According to the expression of $f_0^{(\rm{eq})}$, we have
\begin{equation}\label{f0_eq}
	f_0^{({\rm{eq}})} = \frac{{{\omega _0} - 1}}{{c_s^2}}p + {s_0}\left( {\bf{u}} \right).
\end{equation}
By simply substituting $f_0^{({\rm{eq}})}$ with the expression of the distribution function $f_i$, we can derive the formula for calculating pressure. First, from Eq. (\ref{epsilon_1}) we obtain
\begin{equation}
	\varepsilon f_i^{\left( 1 \right)} =  - {\tau _f}\Delta t \left[{D_i}f_i^{\left( 0 \right)} - \frac{1}{{\Delta t}}f_i^B - {F_i}\right],
\end{equation}
and its equivalent form
\begin{equation}\label{fi_fieq}
	{f_i} - f_i^{({\rm{eq}})} =  - {\tau _f}\Delta t \left[ {D_i}f_i^{\left( 0 \right)} - \frac{1}{{\Delta t}}f_i^B - {F_i}\right].
\end{equation}
Taking the zeroth-direction of Eq. (\ref{fi_fieq}), we have
\begin{equation}
	{f_0} - f_0^{({\rm{eq}})} =  - {\tau _f}\Delta t\left[ {{D_i}f_0^{\left( 0 \right)} - \frac{1}{{\Delta t}}f_0^B - {F_0}} \right].
\end{equation}
Note that the term ${D_i}f_0^{\left( 0 \right)}$ is the order of $O\left( {M{a^2}} \right)$ \cite{Yuan_CMA2020}, then we obtain
\begin{equation}\label{f0_f0_eq}
	{f_0} - f_0^{({\rm{eq}})} = {\tau _f}f_0^B + {\tau _f}\Delta t{F_0} + O\left( {\Delta tM{a^2}} \right).
\end{equation}
Neglecting the term of $O\left( {\Delta tM{a^2}} \right)$, substituting Eq. (\ref{f0_f0_eq}) into Eq. (\ref{f0_eq}), we obtain
\begin{equation}
	\begin{aligned}
		\frac{{{\omega _0} - 1}}{{c_s^2}}p &= {f_0} - {\tau _f}f_0^B - {\tau _f}\Delta t{F_0} - {s_0}\left( {\bf{u}} \right)\\&
		= \sum {{f_i}}  - \sum\limits_{i \ne 0} {{f_i}}  - {\tau _f}f_0^B - {\tau _f}\Delta t{F_0} - {s_0}\left( {\bf{u}} \right)\\&
		=  - \sum\limits_{i \ne 0} {{f_i}}  - {\tau _f}f_0^B - {\tau _f}\Delta t{F_0} - {s_0}\left( {\bf{u}} \right)
	\end{aligned}
\end{equation}
As a result, the calculation of pressure can be given as
\begin{equation}
	p = \frac{{c_s^2}}{{1 - {\omega _0}}}\left[ {\sum\limits_{i \ne 0} {{f_i}}  + {s_0}\left( {\bf{u}} \right) + {\tau _f}\Delta t{F_0} + {\tau _f}f_0^B} \right].
\end{equation}

\section*{Acknowledgments}

J.Y.W. is very grateful to Fang Xiong for the helpful discussions. This work is financially supported by the National Natural Science Foundation of China (Grant No. 12472297).

\end{document}